\documentclass[10pt]{article}
\usepackage{geometry}
\usepackage{cite}
\usepackage{amssymb,amsmath}
\usepackage{wrapfig}
\usepackage{graphicx}
\usepackage[small]{caption}
\usepackage[titletoc,title]{appendix}
\usepackage{multirow}
\usepackage{slashbox}
\usepackage{authblk}

\begin{document} 
\title{Towards a Particle-Number Conserving Theory of Majorana Zero Modes in p+ip Superfluids}
\author[1]{Yiruo Lin \thanks{yiruolin@illinois.edu, linyiruo1@gmail.com}}
\author[1,2]{Anthony J. Leggett \thanks{aleggett@illinois.edu}}
\affil[1]{\textit{Department of Physics, University of Illinois at Urbana-Champaign}}
\affil[2]{\textit{Shanghai Center for Complex Physics, Shanghai Jiaotong University}}

\date{}
\maketitle

\begin{abstract}
Majorana zero modes are predicted to exist in p+ip (either inherent or effective due to proximity effect) superfluids and are proposed to be used for constructing topological qubits for topologically protected quantum computing. Existing theories on the subject are mostly based on BCS mean-field theory which breaks particle number conservation (U(1) symmetry). More specifically, Bogoliubov-de Gennes (BdG) equations are used to derive Majorana zero modes and their braiding properties. The broken particle number conservation is, on the other hand, respected in any fermionic condensed matter system and therefore may be crucial in studying quantum coherence, essential for quantum computing. In this paper, we work towards a particle-number conserving theory by examining the role played by particle number conservation in topological properties of Majorana zero modes. We conclude that the latter may be affected by the former and new theoretical framework that respects particle number  conservation is needed in establishing (or dismissing) the existence of Majorana zero modes in p+ip superfluids and their application to topological quantum computing.
\end{abstract}

\section{Introduction}


The field of quantum information and quantum computing is growing rapidly and is witnessing intriguing interplay with other areas of physics. For our purpose, we'll be focusing on the application of condensed matter systems to quantum computing. We first note a characteristic difference in the two fields: In quantum computing, precise knowledge and accurate control of a particular set of quantum states are usually necessary; whereas in condensed matter physics, we are for most of the time interested in physical quantities that are insensitive to structures (such as topology) of many-body wave functions of the relevant quantum states. The insensitivity is either due to the nature of physical quantities of interest (e.g., local correlation functions) and/or due to the fact that they receive contributions from and are averaged over a macroscopic number of different quantum states for a typical macroscopic condensed matter system. Due to the macroscopic nature of the quantum states of a typical condensed matter system, they are strongly coupled to their environment, resulting in decoherence very quickly, which is disastrous for quantum information processing and quantum computing. Therefore, nontrivial effort is usually required to justify the use of many-body quantum states of condensed matter systems for quantum information and quantum computing. As a well-known example \cite{Leggett}, it was highly nontrivial in the early theoretical developments to understand the behavior of the quantum states under the effect of a dissipative environment, whose interaction with quantum degrees of freedom of interest is complicated, unknown and out of our control. Those developments laid down a theoretical foundation crucial to the later development of superconducting qubit, now a major candidate for quantum computing (for a review, see e.g.,\cite{Clarke}). Despite the great challenges in dealing with decoherence, condensed matter systems have unique advantages over other physical systems thanks to technical advantages in manufacturing and in achieving scalability. \\

Over a little more than the past decade, the notion of topological quantum computing has become very attractive due to its fault-tolerant decoherence-free nature \cite{Kitaev_top_comp}. The whole story hinges on the topological properties of many-body quantum states when condensed matter systems are in a nontrivial topological phase. The subject of topological phases of matter takes central stage on the frontier of condensed matter physics and beyond. As it is somewhat tangential to go into details of this explosively expanding field, we merely note that topological phases are distinct from conventional phases of matter that are characterized by symmetry. Instead, they are characterized by topology of quantum states and hence are immune to local perturbations such as interactions with environments. An early notable condensed matter system exhibiting nontrivial topological properties is 2DEG in semiconducting heterostructure displaying integer quantum Hall effect \cite{IQHE}. The Moore-Read Pfaffian state realized at $\nu=5/2$ quantum Hall plateau in the presence of quasi-holes \cite{MR}, \cite{Chetan} is one of the most prominent candidates in quantum Hall states proposed to hold non-Abelian statistics applicable to topological quantum computing. In this quantum system, the energy levels of the degenerate ground states spanned by these quasi-holes have exponentially small splits which decay with real space separations among the quasi-hole positions and the spectra are separated from the rest by a finite energy gap. No local observations can distinguish between degenerate ground states and the only way to induce nontrivial unitary transformations within the Hilbert space spanned by the degenerate ground state is by braiding the quasi-holes. These properties make the degenerate ground states suitable building blocks for topological quantum qubits. The topological properties of the Moore-Read Pfaffian state with quasi-holes have been carefully justified theoretically \cite{Bonderson}. Following the observation that the Moore-Read Pfaffian state takes the form of BCS pairing with p-wave symmetry \cite{RG}, people have proposed similar topological properties for zero-energy Bogoliubov quasiparticles which are predicted to be localized at Abrikosov vortices in p+ip superfluids \cite{Ivanov}, \cite{salmon},\cite{volv}. These zero-energy quasiparticles are called Majorana zero modes due to their self-adjoint identity and they can be used to build quantum qubits for topological quantum computing. People have subsequently proposed to look for Majorana zero modes in other systems involving effective p-wave pairing terms in the Hamiltonian in certain defects/domain walls between distinct topological phases that can be realized in various hybrid systems involving simple non-topological superfluids (for a review, see e.g., \cite{Alicea_rev}).\\

We first note that the mapping between the Moore-Read Pfaffian state and the superconducting ground state of p+ip superfluids is incomplete. For example, the neutral sector of the former (i.e., the part excluding exponential factors) takes the form of BCS pairing with p-wave symmetry which is not normalizable by itself. In other words, the BCS pairing part of the Pfaffian state doesn't correspond to any physical ground state in superfluids formed in systems such as $^3\mathrm{He}$ in the A phase and $\mathrm{Sr_2RuO_4}$. Furthermore, for the interesting case with Majorana zero modes, there is no explicit mapping available between the Pfaffian state for the $\nu=5/2$ quantum Hall plateau in the presence of quasi-holes and the ground state of p+ip superfluids in the presence of vortices. Therefore, we need to treat physics in the two systems separately. It is worth pointing out that in contrast to the Pfaffian state for $\nu=5/2$ quantum Hall plateau, explicit many-body wave functions in the presence of Majorana zero modes are unavailable to us for physical superfluids (i.e., in contrast to effective superfluids such as formed by composite fermions in quantum Hall systems). It is in this sense and for what we shall discuss next not inappropriate to state that the theoretical basis for physics of Majorana zero modes in topological superfluids is not as sound as that in the quantum Hall case. \\


From now on, we focus on the physics of Majorana zero modes in p+ip superfluids. The name "Majorana zero modes" comes from the observation that each zero-energy Bogoliubov quasiparticle is split into spatially localized parts that are separate in space (the spatial separation suppresses their hybridization exponentially to prevent a energy gap), each of which is self-adjoint, reminiscent of Majorana particles. Due to the presence of zero-energy Bogoliubov quasiparticles, ground states are degenerate. As each zero-energy Bogoliubov quasiparticle is shared by two Majorana zero modes that are spatially separate from each other, information of degenerate ground states is protected from any local probe, in other words, degenerate ground state subspace is protected from local probes. The only way to induce transition among degenerate ground states is by braiding Majorana zero modes that are predicted to obey non-Abelian statistics of Ising type. Furthermore, there is a finite energy gap in the bulk (i.e., superfluid gap), separating the degenerate ground state subspace from the excited eigenstates. The above described properties bear a topological nature in the sense that local perturbations are irrelevant in affecting degenerate ground states, hence we shall refer to them as topological properties. The topological properties of Majorana zero modes make the corresponding degenerate ground state subspace suitable candidate as qubits for quantum computing. In particular, the undetectability of Majorana zero modes ensures that the constructed qubits are immune to local noise and decoherence; the non-Abelian braiding statistics of the former then ensures nontrivial manipulation of the latter. \\

The topological properties of Majorana zero modes described above are derived from the Bogoliubov-de Gennes (BdG) equations. In particular, the self-adjoint identity of a spatially localized zero-energy solution to the BdG equations is due to the particle-hole symmetry of the BdG equations. This symmetry is a consequence of breaking particle number conservation by allowing particle number in the superfluid condensate to fluctuate. We shall discuss the problem associated with breaking particle number conservation in the following paragraph. It is worth noting that the self-adjoint identity is crucial for deducing topological properties of Majorana zero modes. This is related to Fermi statistics which forbids a Majorana zero mode from existing by itself as self-adjoint identity violates the former. Thus one fermionic degree of freedom (which is the fundamental degrees of freedom of excitations in p+ip superfluids) involves at least a pair of Majorana zero modes, indicating the long-range entanglement among spatially separate Majorana zero modes. The long-range entanglement is in turn essential for the non-detectability of Majorana zero modes. Furthermore, both the non-detectability and the braiding statistics depend implicitly on the non-interacting quasiparticle picture as indicated by the BdG equations, in which a system eigenstate is described as non-interacting Bogoliubov quasiparticles (which are superpositions of particle and hole) arising from the (quasiparticle) vacuum and different eigenstates can be related to each other by creating/annihilating the quasiparticles. \\

In the BdG framework, particle number conservation is broken by the mean-field approximation, a standard practice in treating superfluids. It is however not justified and may be even misleading when the same treatment is invoked in analyzing Majorana zero modes in superfluids. The most obvious issue is that the corresponding many-body quantum eigenstates don't have fixed fermion number, but are coherent superposition of different fermion numbers. So they do not represent any physical state of a condensed matter system which allows only fixed fermion number. It is worthing emphasizing here that allowing particle exchange with the environment does not justify the use of quantum superpositions (as distinct from mixtures) of states of different fermion numbers. So to study quantum behavior (which plays the key role in quantum information and quantum computing) of a superfluid system exchanging particles with its environment, we need to take proper account of the universe (superfluid plus environment) that has fixed fermion number. Furthermore, as illustrated in Appendix \ref{sum rules}, since BCS mean-field theory does not take explicit account of the condensate, it can fail in describing physical quantities when the condensate itself has interesting structure (e.g., non s-wave pairing, cf., Appendix \ref{example} or moving condensate, cf. Appendix \ref{Galilean}) and/or when the the condensate plays an explicit role. The failure is already evidenced for the well-known BCS superconducting ground state wave function which violates sum rules related to particle number conservation, and condensate deformation is necessary to account for its zero-point fluctuations (see Appendix \ref{sum rules}). So the key question is whether the true ground states of interest lie in the same manifold as those in the BdG framework in the sense that we can go smoothly from the ground states in the fixed particle number sector constructed in the BdG framework to the true ground states without closing any gap. To address this question, we restrict our attention to the effect of particle number conservation and more specifically to the possible effect of the Cooper pair as a consequence of particle number conservation on the topological properties of degenerate ground states. To get a rough idea of the possible effect from the Cooper pair, it is worth mentioning a plausible argument in favor of neglecting the effect of any extra Cooper pair as total particle number is being fixed. The most straightforward way to recover particle number conservation is to associate a Cooper pair creation operator to the hole component of an operator that creats a Bogoliubov quasiparticle, making the operator particle-number conserving. Since in the BdG framework the extra Cooper pair is extensive spreading over the whole system, its effect is intuitively negligible on the local detectability of the degenerate ground states and also on the Berry phase of braiding Majorana zero modes as long as the involved braiding trajectories cover regions whose sizes are negligible to that of the whole system. Such intuitive argument is unable to address possible finite system effect (which is important in realistic systems) and furthermore ceases to apply even in the thermodynamic limit if there exists deformation of the Cooper pairs near localized Majorana zero modes as consequences of imposing particle number conservation. At this point, it is worth noting that the Cooper pair deformation also invalidates the simple non-interacting quasiparticle picture in the standard BdG framework mentioned in the above paragraph. \\

As discussed in refuting the intuitive argument of neglecting the effect of any extra Cooper pair in the above paragraph, there are two levels of restoring particle number conservation. At the naive level, we restore constant fermion number by projecting the ground states constructed from the BdG equations onto fixed fermion number sector with fermion number matching that of the system (we shall call them projected BdG wave functions in the following). At this level, the effect of particle number conservation is manifest through quasiparticle-associated Cooper pair contribution to Berry phases in braiding Majorana zero modes. We note that  in the standard mean-field theory, the detailed form of Cooper pair wave function of the condensate doesn't play any explicit role in determining the topological properties of Majorana zero modes (instead, it only contributes to the existence of Majorana zero modes via the topology of the corresponding superfluid order parameter). Once fixed fermion number constraint is enforced, the Cooper pair enters explicitly in the Berry phase calculation. At the more sophisticated level, the project BdG wave functions obtained at the above naive level receive further modification as Cooper pairs deform near the locations of Majorana zero modes as consequences of particle number conservation. The Cooper pair deformation can have non-negligible effect on topological properties of Majorana zero modes owing to macroscopic number of Cooper pairs which amplifies $O(1/N)$ modification of each Cooper pair to $O(1)$ corrections.  \\

In this paper, we address the following questions concerning whether taking proper account of particle number conservation spoils the "established" conclusions on the properties of Majorana zero modes (a) about existence of Majorana zero modes (or equivalently ground state degeneracy) (b) about their braiding statistics (c) about their local undetectability. \\


As explicit form of ground state wave functions in general is unknown, we start by considering the simplest situations where analytic solution can be pursued as far as possible. We have considered Berry phase arising from transport of a Bogoliubov quasiparticle around an annulus in which a simple s-wave superfluid with total odd number of fermions is confined \cite{Lin_Leggett_1}. The Cooper pair in the superfluid has quantized winding number around the annulus to simulate a physical vortex in superfluids. External Zeeman field is imposed on a local region to bound the quasiparticle.  We find that superfluid condensate and particle number conservation plays nontrivial role in obtaining the correct Berry phase, particularly when superfluid velocity is nonzero. We have also considered braiding Majorana zero modes in one dimensional Kitaev-like wire network which doesn't exchange fermions with its environment and whose explicit ground state wave functions at the mean-field level can be obtained (see Appendix \ref{Kitaev}). Inspired by these toy models, we then calculate braiding statistics of Majorana zero modes in vortices of p+ip superfluids explicitly taking into account superfluid condensate contribution to Berry phase as a result of fixing total fermion number. Following this, we consider further effect of particle number conservation, specifically effect of modification to superfluid condensate wave function on properties of Majorana zero modes beyond mean-field BdG description. To evaluate such effect beyond the account of the standard framework, we again have to resort to some toy model systems where analytic solutions may be obtained. Our studies suggest possibility of qualitative modifications to physics of Majorana zero modes beyond what predicted by the standard framework. \\

To the best of the authors' knowledge, our work presented here is the first attempt to go beyond the BdG equations to construct a particle-number conserving theory of Majorana physics in p-wave superfluids. The claim needs some clarification. To distinguishing our work from other work that goes beyond or doesn't rely on the BdG equations, we briefly comment on the nature of the existing studies in the literature. Roughly speaking, they can be classified into the following categories: energy spectrum calculations in p-wave superfluids beyond the BdG equations \cite{Park}, \cite{Lopes}; exactly solvable systems that are distinct from p-wave superfluids \cite{Sau} -\cite{Ortiz}; effective p-wave superfluids with fixed particle number but without account of quantum effect from superfluid condensate \cite{Alicea_pc}. None of them covers the effect of particle number conservation on Majorana zero modes in physical p-wave superfluids (as opposed to effective superfluids formed by composite fermions or with fine-tuned pairing interactions for exact solvability) that we consider in this work. \\

The paper is organized as follows. In section \ref{list}, we define our problems and list the main results; in section \ref{braid}, we discuss braiding Majorana zero modes in vortices of p+ip superfluids with fixed total fermion number; in section \ref{condensate deform} and \ref {particle localization}, we evaluate  respectively condensate deformation and net particle localization due to the presence of bound Bogoliubov quasiparticles in superfluids as consequences of particle number conservation beyond the account of the standard mean-field framework; in section \ref{MZM beyond BdG}, we discuss implications of results obtained in section \ref{condensate deform} and \ref{particle localization} to properties of Majorana zero modes; finally in section \ref{summary}, we summarize our results.

\section{Problem setup and main results} \label{list}

We are mainly interested in braiding Majorana zero modes in vortices of 2D p+ip superfluids which do not exchange fermions with their environments. The system of interest is inherently superconducting and has fixed fermion number. The simplest situation to realize a qubit is with four Majorana zero modes and we will be focusing on physics for this case. In order to further simplify as well as to quantify our analysis, we also consider case with two Majorana zero modes. \\

We study effect of particle number conservation in two steps. In the first stage, we study projected BdG wave functions and we found that the Cooper pair contribution to the Berry phase may be non-vanishing if the braiding trajectories enclose areas finite compared to the total system area. The non-vanishing Berry phase contribution results in finite modification to the braiding statistics in the standard framework. The results are summarized in the first row for braiding in table \ref{main results}. \\

To further explore the effect of particle number conservation, we then consider modification of ground state wave functions to the projected BdG  functions as consequences of particle number conservation. At this more sophisticated level, we shall challenge questions (a) and (c) raised above besides the braiding statistics (question (b)). Our analysis suggests possible nontrivial corrections to braiding statistics of Majorana zero modes, including the thermodynamic limit in which dimensionless areas enclosed by braiding trajectories vanish. Furthermore, the very existence of Majorana zero modes becomes questionable as local undetectability and ground state degeneracy may be modified from the mean-field theory. All the modifications we consider to the mean-field theory results arise due to condensate deformation as consequence of condensate and local zero-energy Bogoliubov quasiparticle interplay when particle number conservation is taken into account. This interplay is, roughly speaking,  missing in the mean-field theory as Cooper pair number in the condensate is allowed to fluctuate. The results are summarized in the rest of table \ref{main results}. \\

\begin{table}[h]
\centering
\begin{tabular}{|p{4.3cm}|p{5.5cm}|p{5.5cm}|}
\hline
 \backslashbox{Topological \\ Properties}{Superfluid \\System}
&\makebox{2-Vortex system}&\makebox{ 4-Vortex system } \\
\hline
\multirow{2}{*}{Braiding} & Berry phase has correction$^1$ $O(S)$ & Berry phase may have correction \quad $<O(S)$ \\ \cline{2-3}
& Possible finite$^2$ correction from condensate deformation & possible finite$^2$ correction from condensate deformation \\ \hline
Local Distinguishability & \multicolumn{2}{p{11cm}|}{localized quasiparticles may change local particle density beyond BdG framework} \\ 
\hline
GS Degeneracy & \multicolumn{2}{p{11cm}|}{condensate deformation due to localized quasiparticles may change GS degeneracy, e.g., MZMs become ordinary local modes?} \\ 
\hline
\end{tabular}
\caption{Main results of effect of particle number conservation on properties of Majorana zero modes. 1. $S$ is dimensionless area enclosed by trajectory of interchanging two vortices. 2. finite in the sense that the correction approaches non-zero value in the thermodynamic limit } \label{main results}
\end{table}

\section{Braiding Majorana zero modes in vortices of p+ip superfluids} \label{braid}

In this section, we consider braiding statistics of Majorana zero modes in vortices of p+ip superfluids at the naive level of particle number conservation, i.e., the ground state wave functions are projected BdG wave functions defined above. To facilitate illustrating physics that is distinct from the mean-field theory, we first briefly review the derivation of braiding statistics of Majorana zero modes in the mean-field theory and discuss where in the derivation fixed fermion number may change the final result. \\

Consider a Hamiltonian $H(\lambda)$ parameterized by $\lambda(t)$ varying as a function of time, which admits two instantaneous degenerate ground states for all $t$ as labeled by $|0(\lambda)\rangle$ and $|1(\lambda)\rangle$ respectively. The Hamiltonian goes back to itself after some time interval as $\lambda$ returns to its initial value $\lambda_f=\lambda_i$ with $\lambda_f\equiv\lambda(t_f)$ and $\lambda_i=\lambda(t_i)$. We consider adiabatic evolution of the two ground states $|0(t)\rangle$ and $|1(t)\rangle$ which coincide with the instantaneous ground states at $t_i$, i.e., $|0(t_i)\rangle=|0(\lambda_i)\rangle$ and $|1(t_i)\rangle=|1(\lambda_i)\rangle$. Note that we label the ground states that evolve in time by time variable $t$ and  the instantaneous ground states by $\lambda$. \\

We are interested in the quantum phases picked up by the two ground states at $t_f$ assuming they evolve back to themselves, i.e., $|0(t_f)\rangle=\mathrm{exp}\{i\chi_0\}|0(t_i)\rangle$ and $|1(t_f)\rangle=\mathrm{exp}\{i\chi_1\}|1(t_i)\rangle$. The quantum phases $\chi_0$ and $\chi_1$ can be calculated from knowledge of instantaneous ground states. They consist of two parts: monodromy phase and Berry phase. The former comes from explicit phase of the instantaneous ground states, i.e., $|0(\lambda_f)\rangle=\mathrm{exp}\{i\eta(0)\}|0(\lambda_i)\rangle$ and $|1(\lambda_f)\rangle=\mathrm{exp}\{i\eta(1)\}|1(\lambda_i)\rangle$; and the latter is given by $\phi_0=-\mathrm{Im}\int_{\lambda_i}^{\lambda_f} d\lambda \langle 0(\lambda)|\partial_{\lambda}|0(\lambda)\rangle$ and $\phi_1=-\mathrm{Im}\int_{\lambda_i}^{\lambda_f}d\lambda \langle1(\lambda)|\partial_{\lambda}|1(\lambda)\rangle$. The quantum phase is the sum of the two contributions and the quantity of interest is the relative quantum phase of the two ground states after adiabatic evolution
\begin{eqnarray}
\chi\equiv\chi_1-\chi_0=\delta\eta+\delta\phi, \label{total phase}
\end{eqnarray}
with $\delta\eta=\eta(1)-\eta(0)$ and $\delta\phi=\phi_1-\phi_0$. Note that for simplicity of discussion, we have assumed a simple situation where both monodromy phase and Berry phase are diagonal, i.e., instantaneous ground states at $\lambda_f$ don't mix with each other's initial states at $\lambda_i$ and the off-diagonal Berry's connection vanishes: $\langle0(\lambda)|\partial_\lambda|1(\lambda)\rangle=0$ for all $\lambda$. We shall call such a basis of instantaneous ground states the diagonal basis. The simple situations considered here suffice for our purposes. It's obvious that when $\chi$ is non-zero modulo $2\pi$, a general set of initial ground states will will end up as linear combination of their initial states at the end of the adiabatic process if we choose them to be linear combinations of instantaneous ground states in the diagonal basis. In other words, they will undergo non-trivial unitary evolution in the ground state space. Therefore, non-zero $\chi$ implies non-Abelian statistics when we consider braiding at least three such anyons (in our case, the minimum number of Majorana zero modes is four). \\

It may help now to illustrate through a simple example the concept of the above introduced adiabatic quantum phase, especially that of the monodromy phase. Consider the adiabatic evolution of a spin-1/2 in its ground state in an external magnetic field $\vec{B}$. The Hamiltonian is $H=-\vec{B}\cdot\vec{\sigma}$ and the Schrodinger equation for the spin-1/2 is 
\begin{eqnarray}
\left(\begin{array}{cc}B_z & B(r) \\ B^*(r) & -B_z \end{array}\right)\left(\begin{array}{c}s_\uparrow \\ s_\downarrow\end{array}\right) =-E\left(\begin{array}{c}s_\uparrow \\s_\downarrow\end{array}\right). \label{spin_Schrodinger}
\end{eqnarray} 

\begin{figure}[h!]
\begin{center}
\includegraphics[width=0.3\textwidth]{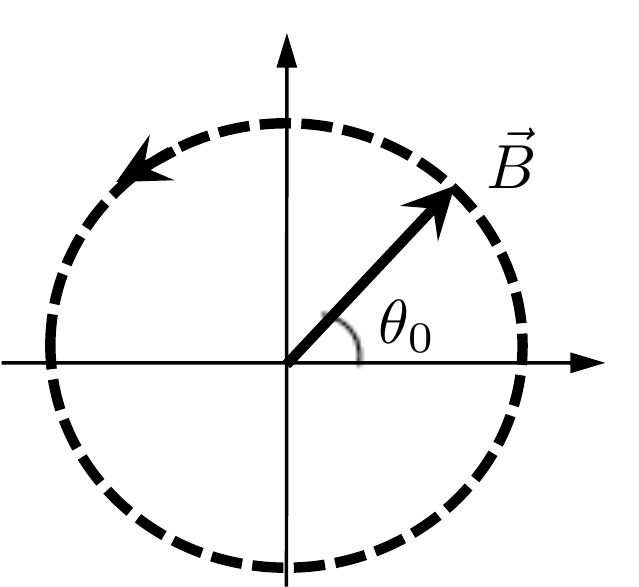}
\end{center}
\caption{A spin-1/2 in a magnetic field whose orientation is parameterized by angle $\theta_0$. The spin is aligned to the direction of the magnetic field in its ground state. The arrow on the dashed circle indicates the direction of rotating the magnetic field $\vec{B}$.}
\label{fig_spin}
\end{figure}

For simplicity, we consider the situation where the magnetic field is in the plane, i.e., $B_z=0$ (cf. figure \ref{fig_spin}). The Schrodinger equation now reads
\begin{eqnarray}
\left(\begin{array}{cc}0 & \mathrm{exp}\{i\theta_0\} \\ \mathrm{exp}\{-i\theta_0\} & 0 \end{array}\right)\left(\begin{array}{c}s_\uparrow \\ s_\downarrow\end{array}\right) =-E\left(\begin{array}{c}s_\uparrow \\s_\downarrow\end{array}\right). \label{spin_Schrodinger_1}
\end{eqnarray} 
We calculate the ground state quantum phase as we adiabatically rotate the magnetic field by $2\pi$. Both the monodromy phase and the Berry phase depend on the choice of the instantaneous ground state. If we choose the instantaneous ground state $|0(\theta_0)\rangle$ to be single-valued function of $\theta_0$, it can be chosen to be $\frac{1}{\sqrt{2}}(\mathrm{exp}\{i\theta_0\},1)^T$. The monodromy phase in this gauge is zero by definition and the Berry phase is calculated to be
\begin{eqnarray}
\phi&=&-\mathrm{Im}\{\int_0^{2\pi}\langle0(\theta_0)|\partial\theta_0|0(\theta_0)\rangle\} \nonumber \\
&=&-\pi, \label{Berry}
\end{eqnarray}
So the total quantum phase in the adiabatic evolution is equal to the Berry phase, i.e., $-\pi$. \\

If we instead choose the instantaneous ground state to be, say, $\frac{1}{\sqrt{2}}(\mathrm{exp}\{i\theta_0/2\},(\mathrm{exp}\{-i\theta_0/2\})^T$, i.e., we multiply the above chosen instantaneous ground state by $\mathrm{exp}\{-i\theta_0/2\}$, then the instantaneous ground state picks up a minus sign after the magnetic field is rotated by $2\pi$, so the monodromy phase is $\pi$. The Berry phase is calculated to be zero as the contributions from spin-up and spin-down component cancel. The total quantum phase is equal to the monodromy phase $\pi$, which is the same as the above result confirming that the adiabatic quantum phase is gauge independent. \\

\begin{figure}[h!]
\begin{center}
\includegraphics[width=0.45\textwidth]{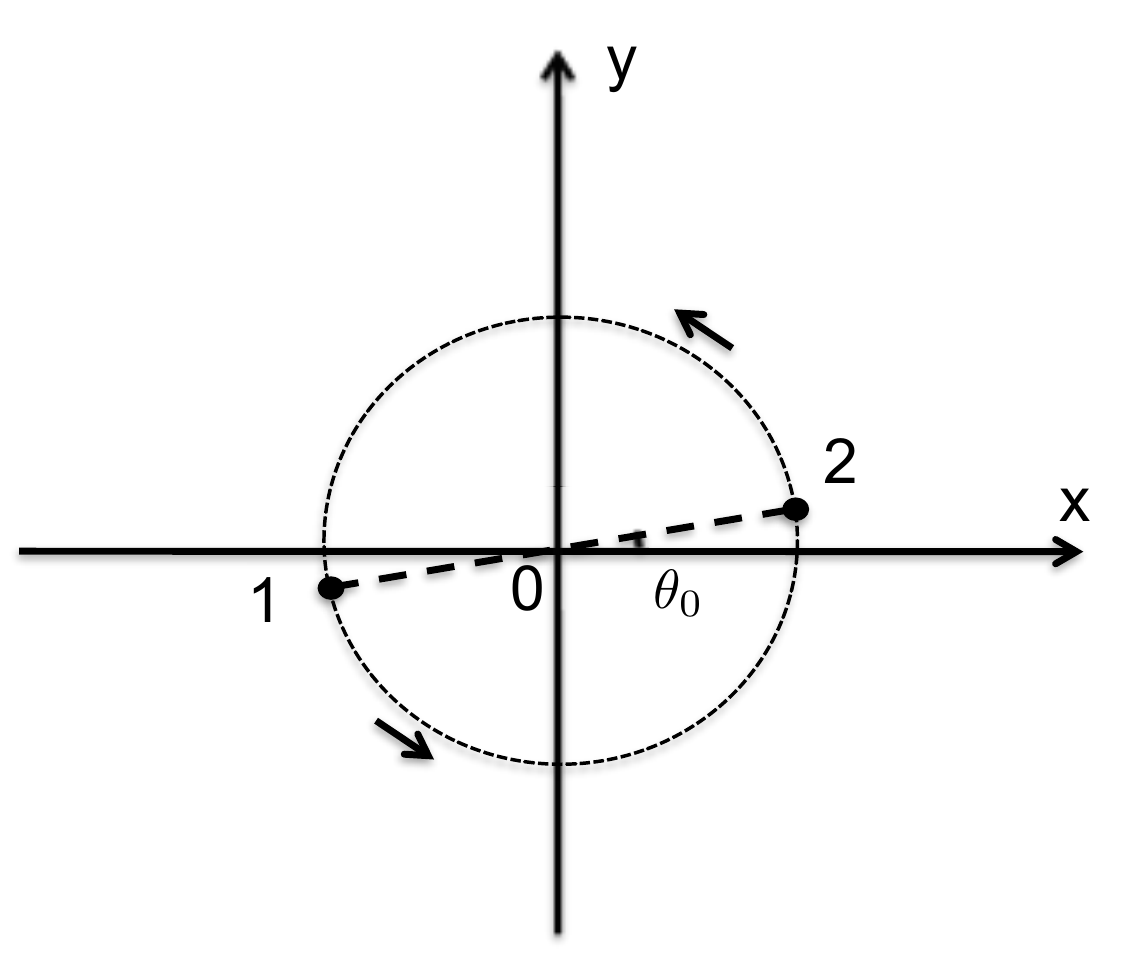}
\end{center}
\vspace{-15pt}
\caption{Configuration of braiding two vortices. The dashed circle centered around $0$ with arrows indicates the trajectories of vortices 1 and 2 during the counter-clockwise exchange. $\theta_0$ parameterizes the position of the two vortices.}
\label{fig_vortex2}
\end{figure}

We now first specialize to the system of a 2D p+ip superfluid containing a pair of vortices, each harboring a Majorana zero mode at the respective vortex core. At the mean-field level, the system has double degenerate ground states which differ in fermion number parity and are connected by the pair of Majorana zero modes. We choose the instantaneous ground states as 
\begin{eqnarray}
|1(\theta_0)\rangle=\alpha^\dagger_0|0(\theta_0)\rangle, \label{0_1}
\end{eqnarray}
where $\alpha^\dagger_0$ is the zero-energy Bogoliubov quasiparticle creation operator which is formed out of the pair of Majorana zero modes
\begin{eqnarray}
\alpha^\dagger_0=\frac{1}{2}(\gamma_1+i\gamma_2), \label{alpha_0}
\end{eqnarray}
where $\gamma_1$ and $\gamma_2$ are two Majorana zero modes sitting at vortex 1 and 2 respectively, given by
\begin{eqnarray}
\gamma_i(\theta_0)=\int d^2ru_i(r,\theta_0)\psi^\dagger(r)+v_i(r,\theta_0)\psi(r), \label{gamma_pnc}
\end{eqnarray}
where $i=1,2$ and $u_i(r,\theta_0)=v^*_i(r,\theta_0)$ are functions localized at vortex $i$. The normalization condition for $u_i(r,\theta_0)$ is $\int d^2r\hspace{3pt} |u_i(r,\theta_0)|^2=1$. $\theta_0$ is a parameter characterizing the position of the vortex pair (see figure \ref{fig_vortex2}) with initial value $0$ and final value $\pi$ before and after interchanging the two Majorana zero modes, respectively. \\

To facilitate the following discussion of interchanging two Majorana zero modes, it is helpful to first describe the topological property of the superfluid order parameter and of the Cooper pair in the presence of a pair of vortices. Note that we don't consider the case of a vortex-antivortex pair because their interchange does not lead back to the same state. The quantum phases of the superfluid order parameter and of the Cooper pair have quantized winding number. They get two contributions: the intrinsic Cooper pair angular momentum of unit quantum number due to p+ip pairing, and the center-of-mass angular momentum due to the pair of vortices. Each vortex produces one quantum of angular momentum with respect to its vortex core (valid for infinite systems; for finite systems, see discussions below). Due to the localized nature of Majorana zero modes, the local angular momentum of the order parameter relevant for each Majorana zero mode is two (assuming the same angular momentum due to each vortex and the Cooper pair throughout the discussion). The contribution from the other vortex vanishes when evaluated near the vortex core in consideration and the net effect is an overall phase. For the Cooper pair, since it spreads over the whole system, its total angular momentum is the sum of the intrinsic one and the center-of-mass one from both vortices, and hence equals three. \\

Consider now interchanging vortex 2 with vortex 1, as shown in figure \ref{fig_vortex2}.  Since after the interchange, the system Hamiltonian goes back to the initial one, the instantaneous ground states $|0(\theta_0)\rangle$ and $|1(\theta_0)\rangle$ must also go back to their initial states up to monodromy phases. In the adiabatic limit, each ground state will evolve back to itself at the end of the interchange up to overall phases. They wouldn't evolve into linear superposition of the two basis ground states since the latter have different particle number parity. For the more relevant case of four Majorana zero modes, see below.  \\

The Majorana wave functions $u_i(r,\theta_0)$ can be written as \cite{Gurarie}
\begin{eqnarray}
u_1(r,\theta_0)&=&\mathrm{exp}\{\frac{(\pi+\theta_0) i}{2}\}u(|\vec{r}-\vec{R}_1|)e^{i\theta(\vec{r}-\vec{R}_1)} \nonumber \\
u_2(r,\theta_0)&=&\mathrm{exp}\{\frac{\theta_0 i}{2}\}u(|\vec{r}-\vec{R}_2|)e^{i\theta(\vec{r}-\vec{R}_2)}, \label{u_case1}
\end{eqnarray}
as zero-energy solutions to the BdG equations for a p+ip superfluid around a vortex with unit winding number, where $\theta(\vec{r}-\vec{R_i})$ is polar angle of vector $\vec{r}-\vec{R_i}$, $\vec{R}_i$ is the location of vortex i (i=1,2). The overall $\theta_0$-dependent phases for $u_1$ and $u_2$ come from overall phases of gap function at each vortex core $R_i(i=1,2)$ due to the superfluid phases induced by the other vortex. The superfluid phase increases by $2\pi$ going around each vortex counter-clock-wisely. The overall superfluid phase at vortex 1 is $\pi+\theta_0$ and $\theta_0$ at vortex 2 (see figure \ref{fig_vortex2} for vortex configuration, we define zero phase value in the positive x direction). So the overall phases for $u_1$ and $u_2$ are $(\pi+\theta_0)/2$ and $\theta_0/2$, respectively (see also \cite{Ivanov} for essentially the same point on the overall phases for $u_1$ and $u_2$). The azimuthal dependence of $u_1$ and $u_2$ with respect to the corresponding vortex core is given by phases of azimuthal angles $\theta(\vec{r}-\vec{R}_1)$ and $\theta(\vec{r}-\vec{R}_2)$ since $u_1$ and $u_2$ are eigenstates of angular momentum (with eigenvalue one) with respect to their associated vortices.  \\

After the interchange, $\theta_0$ goes from 0 to $\pi$, the Majorana wave functions become, by (\ref{u_case1}) 
\begin{eqnarray}
u_1(r,\pi) &=& -u_2(r,0), \nonumber \\
u_2(r,\pi)&=&u_1(r,0), \nonumber 
\end{eqnarray}
so that 
\begin{eqnarray}
\alpha^\dagger_0(\pi)&=&i\alpha^\dagger_0(0). \label{alpha_pi}
\end{eqnarray}

Combining equation (\ref{alpha_pi}) with the definition of the instantaneous ground states (\ref{0_1})-(\ref{gamma_pnc}),  we obtain the monodromy phase
\begin{eqnarray}
\delta \eta=\pi/2. \label{delta_alpha}
\end{eqnarray}

Now let's calculate the Berry phase. It can be written as follows
\begin{eqnarray}
\delta \phi&=&-\mathrm{Im}\{\int^{\pi}_0 d\theta_0 \langle 1(\theta_0)|\partial_{\theta_0}|1(\theta_0)\rangle-\int^{\pi}_0 d\theta_0 \langle 0(\theta_0)|\partial_{\theta_0}|0(\theta_0)\rangle\} \nonumber \\
&=&-\mathrm{Im}\{\int^{\pi}_0 d\theta_0 \langle 0(\theta_0)|\alpha(\theta_0)[\partial_{\theta_0},\alpha^\dagger(\theta_0)]|0(\theta_0)\rangle\} \nonumber \\
&=&-\mathrm{Im}\{\int^{\pi}_0 d\theta_0 \langle 0(\theta_0)|\alpha(\theta_0)(\partial_{\theta_0}\alpha^\dagger(\theta_0))|0(\theta_0)\rangle\} \nonumber \\
&=&-\mathrm{Im}\{\int^{\pi}_0 d\theta_0 \langle 0(\theta_0)|\{\alpha(\theta_0),(\partial_{\theta_0}\alpha^\dagger(\theta_0))\}|0(\theta_0)\rangle\}.\label{Berry_0_1}
\end{eqnarray}
In obtaining the above equation, $|1(\theta_0)\rangle=\alpha^\dagger(\theta_0)|0(\theta_0)\rangle$ and $\alpha(\theta_0)\alpha^\dagger(\theta_0)|0(\theta_0)\rangle=|0(\theta_0)\rangle$ have been used to go from the first equality to the second, and $\alpha(\theta_0)|0(\theta_0)\rangle=0$ has been used to derive the last equality. It's now straightforward to show that the contribution from derivatives of $u_i(r,\theta_0)$ and $v_i(r,\theta_0)$ vanishes from the last equality in the above equation and so $\delta\phi=0$. The vanishing contribution is due to two characteristics of localized Majorana zero modes: reality condition of $\gamma_i$ so that $u_i=v_i^*$; $u_1$ and $u_2$ are localized at separate regions so their overlap is exponentially small. More specifically, the anti-commutator of the last equality can be evaluated as follows by substituting equations (\ref{alpha_0}) and (\ref{gamma_pnc}) into $\alpha(\theta_0)$ and $\alpha^\dagger(\theta_0)$
\begin{eqnarray}
\{\alpha(\theta_0),(\partial_{\theta_0}\alpha^\dagger(\theta_0))\}&=&\int d^2r \partial_{\theta_0}(|u_1(r,\theta_0)|^2+|u_2(r,\theta_0)|^2) \nonumber \\
&+& i\int d^2r(u^*_1(r,\theta_0)\partial_{\theta_0}u_2(r,\theta_0)-u^*_2(r,\theta_0)\partial_{\theta_0}u_1(r,\theta_0)+c.c.) \nonumber \\
&=& 0 \nonumber 
\end{eqnarray}
As we will see next, once Cooper pair operator is added to the BdG quasiparticle operator, $\gamma_i$ are no longer self-adjoint as they involve the Cooper pair breaking the particle-hole symmetry. Furthermore, there can be contribution to the Berry phase from the Cooper pair as its form varies during the braiding process. \\

The vanishing relative Berry phase combined with the monodromy phase given by equation (\ref{delta_alpha}) yields the total relative quantum phase
\begin{eqnarray}
\chi=\pi/2+0=\pi/2. \label{phase_Ivanov}
\end{eqnarray}
This is the standard result first derived by Ivanov \cite{Ivanov}. After interchanging vortex 1 with 2, the two ground states picks up different overall quantum phases and the phase difference is $\pi/2$. The contribution comes only from the monodromy phase.\\

Now let's examine as we project the ground states defined by (\ref{0_1})-(\ref{gamma_pnc}) onto fixed fermion number sector, where in the above derivation of $\chi$ modification may occur. Each of the two degenerate ground states can be written in fermion number space as
\begin{eqnarray}
|\Psi(\lambda)\rangle_{\mathrm{BdG}}=\sum_n\beta_n(\lambda)|\Psi_n(\lambda)\rangle \label{BdG_number_space}
\end{eqnarray}
where $n$ label the total fermion number which is either all even or all odd. We choose all coefficients $\beta_n(\lambda)$ to be real and all $|\Psi(\lambda)\rangle)_n$ to be normalized. We have added subscript 'BdG' to the ground states in the mean-field theory to emphasize that they are constructed from the BdG equations. Note that in the above equation, we label the ground states by $|\Psi(\lambda)\rangle$ for a general discussion and we will switch back to the notation for the ground states used in equation (\ref{0_1}) as we come to discussing interchanging two vortices.\\

In the diagonal basis, each instantaneous ground state picks up a monodromy phase at the end of the adiabatic evolution
\begin{eqnarray}
|\Psi(\lambda_f)\rangle_{\mathrm{BdG}}=\mathrm{exp}\{i\alpha\}|\Psi(\lambda_i)\rangle_{\mathrm{BdG}}. \label{mono_BdG}
\end{eqnarray}
Comparison of equation (\ref{mono_BdG}) with (\ref{BdG_number_space}) immediately yields the conclusion that monodromy phase for each state $|\Psi_n(\lambda)\rangle$ with fixed fermion number has to be the same as $\alpha$. Therefore, projecting the BdG states onto fixed fermion number sector wouldn't change the monodromy phase. \\

On the other hand, there is no guarantee for Berry phase to stay unchanged under particle number projection. The Berry phase for each BdG ground state can be written in terms of contributions from states in fermion number space as
\begin{eqnarray}
\phi_{\mathrm{BdG}}=-\sum_n \mathrm{Im}\{\int_{\lambda_i}^{\lambda_f} \hspace{-2pt}d\lambda\hspace{5pt} \beta^2_n(\lambda) \hspace{2pt}\hspace{-1pt}\langle\Psi_n(\lambda)|\partial_\lambda|\Psi_n(\lambda)\rangle\}, \label{Berry_number_space}
\end{eqnarray}
which can be regarded as averaged over Berry phase $\phi_n=-\mathrm{Im}\{\int_{\lambda_i}^{\lambda_f} \hspace{-2pt}d\lambda\hspace{2pt}\hspace{-1pt}\langle\Psi_n(\lambda)|\partial_\lambda|\Psi_n(\lambda)\rangle\}$ for each $|\Psi_n(\lambda)\rangle$. In particular, there is no justification for $\phi_{\mathrm{BdG}}=\phi_{n_0}$ for fermion number equal to that of particle number-conserving system of interest $n=n_0$. Therefore, we may expect correction to the Berry phase resulting from particle number projection.  \\

\begin{figure}[h!]
\begin{center}
\includegraphics[width=0.45\textwidth]{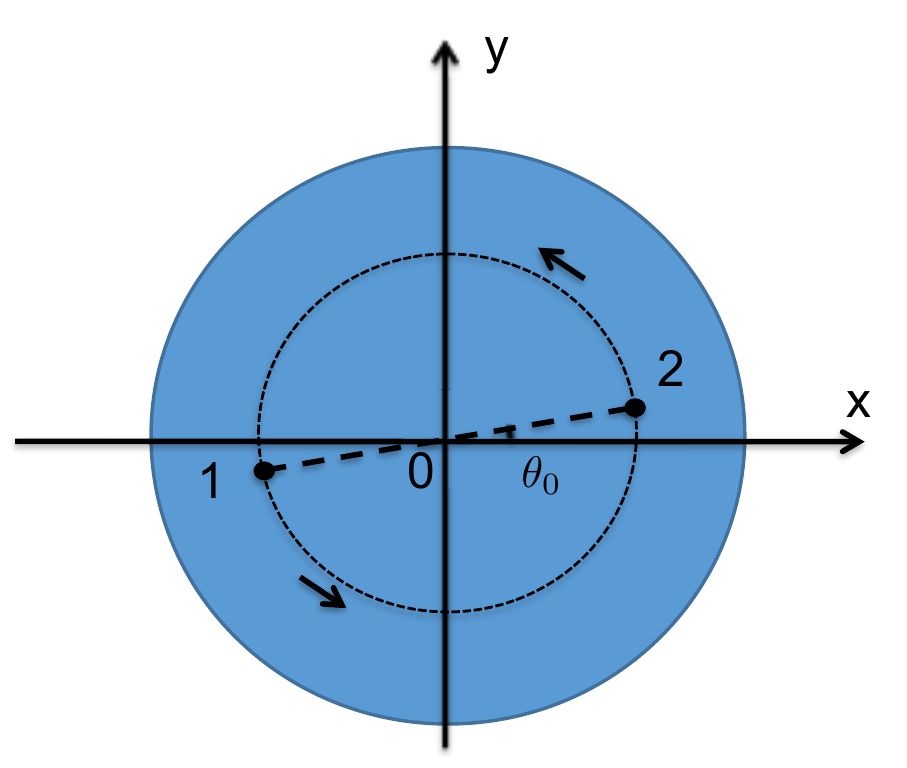}
\end{center}
\vspace{-15pt}
\caption{Configuration of braiding two vortices in a finite system. The dashed circle centered around $0$ with arrows indicates the trajectories of vortices 1 and 2 during the counter-clockwise exchange. $\theta_0$ parameterizes the position of the two vortices. The system occupies a disc region. }
\label{fig_vortex2_finite}
\end{figure}

We next specialize to the Berry phase of interchanging two vortices. To simplify our considerations, we choose the system boundary to be circular, i.e, the 2D p+ip superfluid occupies a disc region (see figure \ref{fig_vortex2_finite}). The relevant quantity is the relative Berry phase of the two projected ground states which, in the initial configuration where $\theta_0=0$, are identical to the projected instantaneous ground states $|0(\theta_0)\rangle_{2N}$ and $|1(\theta_0)\rangle_{2N+1}$ with fermion number $2N$ and $2N+1$ respectively. To relate the two projected instantaneous ground states, we associate explicitly a Cooper pair operator to the zero-energy Bogoliubov quasiparticle operator $\alpha^\dagger_0(\theta_0)$ and label the modified one as $\bar{\alpha}^\dagger_0(\theta_0)\equiv1/2(\bar{\gamma}_1+i\bar{\gamma}_2)$. The modified Majorana zero mode operators now contain Cooper pair operator $C^\dagger(\theta_0)$ which adds a Cooper pair to the system and connects states in different fermion number sectors, i.e., they take the following form 
\begin{eqnarray}
\bar{\gamma}_i(\theta_0)=\int d^2r \hspace {5pt} u_i(r,\theta_0)\psi^\dagger(r)+v_i(r,\theta_0)\psi(r)C^\dagger(\theta_0) \label{gamma_pc}
\end{eqnarray}
with
\begin{eqnarray}
C^\dagger(\theta_0)|0(\theta_0)\rangle_{2N}=|0(\theta_0)\rangle_{2N+2}.  \label{Cooper pair_particle number}
\end{eqnarray}

Compared to the particle-number non-conserving Majorana zero mode operators defined above in equation (\ref{gamma_pnc}), the particle-number conserving version (equation (\ref{gamma_pc})) take the same form except the extra Cooper pair operator. Note that as the two projected ground states have different fermion number, they are no longer degenerate but differ in energy by the chemical potential. Nevertheless, the relevant physics here is independent of the energy splitting due to different fermion number. Note further that the BdG formalism puts a constraint on the form of the Cooper pair operator. To satisfy the requirement that each instantaneous ground state returns to its initial form (up to a monodromy phase, cf. equation (\ref{mono_BdG})) at the end of the adiabatic evolution,  each state in fixed particle number sector (i.e., $|\Psi_n(\lambda)\rangle$ in equation (\ref{BdG_number_space})) must returns to its initial form as well. As the Cooper pair operator connects states in different fermion number sectors (cf. equation(\ref{Cooper pair_particle number})), it must also returns to its initial from. In particular, in the present discussion, we require $C^\dagger(\pi)=C^\dagger(0)$.  \\


The modification to the Berry phase due to the Cooper pair operator can be calculated by equation (\ref{Berry_0_1}) with the particle-number conserving Bogoliubov quasiparticle operator $\bar{\alpha}^\dagger(\theta_0)$ where the derivative is to be taken on the Cooper pair operator $C^\dagger(\theta_0)$,
\begin{eqnarray}
\delta\phi=-\mathrm{Im}\int_0^{\pi} d\theta_0  \hspace{2pt}_{2N}\langle0(\theta_0)|\{\bar{\alpha}_0(\theta_0),\partial_{\theta_0}\bar{\alpha}_0^\dagger(\theta_0)\}|0(\theta_0)\rangle_{2N}. \label{Berry_Cooper}
\end{eqnarray}
The above equation needs some justification. The particle-number conserving Bogoliubov quasiparticle operator $\bar{\alpha}(\theta_0)$ doesn't annihilate projected ground state $|0(\theta_0)\rangle_{2N}$ exactly, inducing correction to the above expression. Since the induced correction is of order $1/N$, we neglect it here and consider only corrections that are finite in the limit $N\rightarrow \infty$. \\

In general, the analytical form of $C^\dagger(\theta_0)$ is unavailable in the presence of vortices. However, we may still estimate its contribution to the Berry phase. The simplest situation is when the area enclosed by the dashed circular trajectory (see figure \ref{fig_vortex2_finite}) as traced by interchanging the two vortices approaches zero, compared to the total system area. In this limit, the dependence of the Cooper pair wave function  
on $\theta_0$ vanishes since the positions of the two vortices are essentially unchanged as viewed from most part of the system. Since the magnitude of the Cooper pair wave function is essentially uniform over the whole region of the system, it is essentially unchanged during the interchange. Therefore, in this limit, particle number projection doesn't change the Berry phase and the braiding statistics of Majorana zero modes stays unchanged from particle-number non-conserving framework. For comparison to an equivalent braiding process, we write down the ansatz for the Cooper pair wave function in this limit
\begin{eqnarray}
C^\dagger(\theta_0)=\int dRdr\hspace{5pt} \mathrm{exp}(2i\Theta+i\theta)\Psi_c(|R|)\Psi_r(|r|) \psi^\dagger(r_a)\psi^\dagger(r_b)\label{Cooper_case1}
\end{eqnarray}
where $R$ and $r$ are center-of-mass and relative coordinates of the two particles (located at $r_a$ and $r_b$, respectively) forming the Cooper pair. $\Theta$  and $\theta$ are center-of-mass and relative polar angles, respectively. $\Psi_c$ and $\Psi_r$ denote functions of center-of-mass and relative coordinates, respectively. They are functions of only magnitudes of $R$ and $r$. The factor of two for $\Theta$ in the exponent comes from the total vorticity due to the vortex pair. \\
 
When the area enclosed by the trajectory of interchanging the two vortices is nonzero ( i.e., non-negligible compared to the total system area), we can no longer neglect the dependence on the vortex configuration of the Cooper pair wave function. In this case, the derivative of the Cooper pair wave function with respect to $\theta_0$ is non-vanishing, from which we may expect finite correction to the Berry phase. In particular, given the vortex-induced non-zero Cooper pair phase winding from the region inside the interchange trajectory during the interchange process, we expect contribution to the Berry phase to come from the inside region (the region inside the dashed circle in figure \ref{fig_vortex2_finite}). Assuming the only contribution to the Berry phase is coming from the vortex-induced Cooper pair phase winding, we may estimate the amount of correction as follows. Since one Cooper pair contributes $-2\pi$ to the Berry phase if it were entirely inside the interchange trajectory \cite{Ao}, the correction to the Berry phase is scaled by the fraction of the Cooper pair inside the interchange trajectory multiplied by another factor of one-half. The additional factor of one-half comes from taking into account that the Cooper pair is associated only with the hole component of the Bogoliubov quasiparticle (cf. equation (\ref{gamma_pc})) and for a bound quasiparticle, the hole component has the same weight as the particle component. So we get an estimate of the correction to the Berry phase to be $-S\pi$, where $S$ is the dimensionless area enclosed by the interchange trajectory. Combined with the monodromy phase, we see that the total phase $\chi$ (defined in equation (\ref{total phase})) receives a correction from particle number projection that is proportional to the area of the interchange trajectory. The correction to particle-number non-conserving result is due to the fact that the two projected ground states differ in total fermion number by one. The extra fermion involves one Cooper pair which contributes to the Berry phase due to its non-trivial phase winding induced by moving the vortices. \\

\begin{figure}[h!]
\begin{center}
\includegraphics[width=0.45\textwidth]{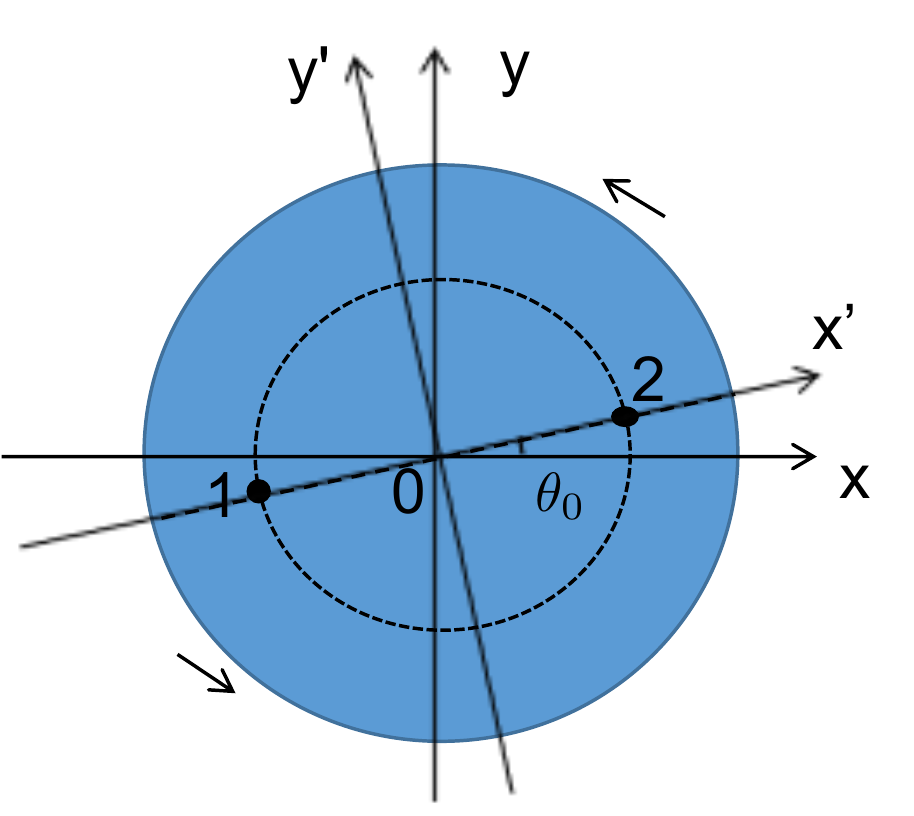}
\end{center}
\vspace{-15pt}
\caption{Configuration of rotating the whole system around its center. The system instantaneous eigenstates are unchanged in the rotated framed labeled with prime superscript. The system occupies a disc region. }
\label{fig_vortex2_rotate}
\end{figure}

The estimated correction to the braiding statistics finds a very nice physical explanation in an equivalent interchanging process in which we rotate the whole system around the origin by $\pi$ (shown in figure \ref{fig_vortex2_rotate}). It is interesting to note, as we shall see, that in this alternative process, the instantaneous ground state wave functions in the particle-number non-conserving framework do NOT return to their initial states at the end of the process. On the other hand, the number projected ground states do return to their initial states. The monodromy phase and the Berry phase are redistributed compared to the standard interchanging process in which only the two vortices are moved. Without taking into account the Cooper pair contribution, the braiding statistics in the alternative process would differ from the standard process. In this sense, Cooper pair plays an essential role in determining the braiding statistics in the alternative process which fails the standard particle-number non-conserving framework.\\

The Majorana wave functions $u_i(r,\theta_0)$ take the following form (Appendix \ref{proof})
\begin{eqnarray}
u_1(r,\theta_0)&=&\mathrm{exp}\{(\frac{\pi}{2}-\theta_0) i\}u(|\vec{r}-\vec{R}_1|)e^{i\theta(\vec{r}-\vec{R}_1)} \nonumber \\
u_2(r,\theta_0)&=&\mathrm{exp}\{-\theta_0 i\}u(|\vec{r}-\vec{R}_2|)e^{i\theta(\vec{r}-\vec{R}_2)}. \label{u_case2}
\end{eqnarray}

After $\pi$ rotation, the BdG operator $\bar{\alpha}^\dagger(\theta_0)$ becomes
\begin{eqnarray}
\bar{\alpha}^\dagger(\pi)=-\bar{\alpha}^\dagger(0) \label{alpha_pi_case2}
\end{eqnarray}
provided that the Cooper pair wave function becomes
\begin{eqnarray}
C^\dagger(\pi)=-C^\dagger(0), \label{C_pi_case2}
\end{eqnarray}
which will be justified shortly. Note that since the Cooper pair wave function changes sign at the end of the rotation, the particle number non-conserving ground states, as coherent superposition of states with different fermion numbers, do not return to their initial states (cf. definition of the Cooper pair operator in equation (\ref{Cooper pair_particle number})). \\

Equation (\ref{alpha_pi_case2}) together with the definition of the two degenerate ground states yield the monodromy phase
\begin{eqnarray}
\delta \eta=\pi, \label{delta_alpha_case2}
\end{eqnarray}
which differs from the value $\pi/2$ in the standard interchanging process.\\

Now, let's calculate the Berry phase. Again, the contribution from the derivatives of $u_i(r,\theta_0)$ and $v_i(r,\theta_0)$ vanishes. So we focus on the contribution from the Cooper pair. In the rotation process, both the relative and center-of-mass coordinates are rotated, so the Cooper pair wave function is dependent on $\theta_i-\theta_0$ for each particle $i$ forming the Cooper pair. In the limit $S\rightarrow 0$ (i.e., dimensionless area enclosed by the trajectory traced by the two vortices compared to the total system area), we have the same ansatz for the Cooper pair as (\ref{Cooper_case1}) with polar coordinates replaced by relative ones, i.e., $\theta_i\rightarrow\theta_i-\theta_0$, (cf. also discussion in Appendix \ref{equiv})
\begin{eqnarray}
C^\dagger(\{\theta_i-\theta_0\})=\int dRdr\hspace{5pt} \mathrm{exp}(2i(\Theta-\theta_0)+i(\theta-\theta_0))\Psi_c(|R|)\Psi_r(|r|) \psi^\dagger(r_a)\psi^\dagger(r_b). \label{Cooper_case2}
\end{eqnarray}

Since $C^\dagger(\{\theta_i-\theta_0\})$ as given by (\ref{Cooper_case2}) is eigenstate of $\partial_{\theta_0}$ with eigenvalue $-3i$, its contribution to the Berry phase is $3\pi/2$ (which can be evaluated straightforwardly by equation (\ref{Berry_Cooper}) keeping in mind again that the Cooper pair is associated with the hole component of the Bogoliubov quasiparticle operator which takes half of the total quasiparticle weight). Together with the monodromy phase of $\pi$, the total braiding phase is
\begin{eqnarray}
\chi=\delta\eta+\delta\phi=\pi+3\pi/2=5\pi/2. \label{phase_case2}
\end{eqnarray}
This is equivalent to $\pi/2$ found in the standard interchanging process. \\

The above calculation of the Cooper pair contribution to the Berry phase can be understood more intuitively from angular momentum (see also \cite{Lin_Leggett_1}). The Berry phase due to the Cooper pair can be regarded as $\pi$ times its average angular momentum (due to the dependence of the Cooper pair on $\theta_0$ when system boundary is circular as assumed here, cf. equation (\ref{Cooper_case2})). In the thermodynamic limit (i.e., $S\rightarrow0$), the particle number projected ground states $|0(\theta_0)\rangle_{2N}$ (and also $|0(\theta_0\rangle_{2N+2}$) can be regarded as possessing rotation symmetry since both vortices are located at the origin as viewed from large distances. So the Cooper pair wave function can be regarded as eigenstate of angular momentum with eigenvalue $3$. Taking into account the weight of the Cooper pair associated with the quasiparticle, its contribution to the Berry phase is $3\pi/2$. \\

For a finite system with a circular boundary, we can make similar estimation as we did for the standard interchanging process. In the alternative process,  the vortex contribution to the Berry phase for the region inside the interchange trajectory vanishes since it corresponds to center-of-mass angular momentum whose average is zero inside the trajectory of the two vortices, thereby decreasing the Berry phase from the thermodynamic limit. Recall that in the standard process, it is the vortex contribution from the same region that contributes to the Berry phase. Hence, the corrections in the two processes are identical owing to an extra sign difference in the two processes, i.e, angular momentum is derivative to $\theta_i$ of each particle forming the Cooper pair which is opposite to derivative to $\theta_0$. The two interchange processes are physically equivalent and the verification is provided in Appendix \ref{equiv}. \\

\begin{figure}[h!]
\begin{center}
\includegraphics[width=0.45\textwidth]{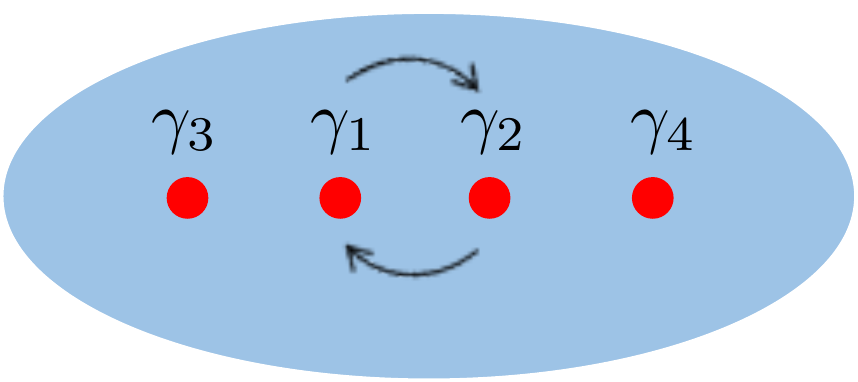}
\end{center}
\vspace{-15pt}
\caption{Configuration of braiding two vortices in a 4-vortex superfluid. Four Majorana zero modes reside in the four vortices and we interchange vortex 1 and 2 while keeping 3 and 4 unchanged.} 
\label{fig_vortex4}
\end{figure}

We now turn to the more relevant case of four Majorana zero modes sitting in four vortices. Since the relevant basis ground states have same particle-number parity, we need to calculate the matrix of Berry connection. Nevertheless, we can always choose our initial ground states to coincide with the instantaneous ground states in the diagonal basis where we only need to consider monodromy phase and Berry phase as described above. We interchange vortex 1 and 2 while fixing the other two (see figure \ref{fig_vortex4}). To project from particle-number non-conserving ground states onto particle-number conserving ground states, we again include explicitly Cooper pair operators. The corresponding projected instantaneous ground states can be related to each other by the following particle-number conserving zero-energy Bogoliubov quasiparticles supplemented with an extra Cooper pair operator to ensure the same fermion number for the doubly degenerate ground states
\begin{eqnarray}
|11(\lambda)\rangle_{2N}=C(\lambda)\bar{\alpha}^\dagger_0\bar{\alpha}^\dagger_1|00(\lambda)\rangle_{2N} \label{00_11}
\end{eqnarray}
with particle-number conserving zero-energy Bogoliubov quasiparticle operators given by
\begin{eqnarray}
\bar{\alpha}^\dagger_0=1/2(\bar{\gamma}_1+i\bar{\gamma}_2) \label{alpha_12}
\end{eqnarray}
and 
\begin{eqnarray}
\bar{\alpha}^\dagger_1=1/2(\bar{\gamma}_3+i\bar{\gamma}_4), \label{alpha_34}
\end{eqnarray}
where particle-number conserving Majorana zero modes are given by
\begin{eqnarray}
\bar{\gamma}_i(\lambda)=\int d^2r \hspace {5pt} u_i(r,\lambda)\psi^\dagger(r)+v_i(r,\lambda)\psi(r)C_{12}^\dagger(\lambda) \label{gamma_12}
\end{eqnarray}
for vortex $i=1,2$
and
\begin{eqnarray}
\bar{\gamma}_i(\lambda)=\int d^2r \hspace {5pt} u_i(r,\lambda)\psi^\dagger(r)+v_i(r,\lambda)\psi(r)C_{34}^\dagger(\lambda) \label{gamma_34}
\end{eqnarray}
for vortex $i=3,4$.
Note that we have switched to notation $\lambda$ parameterizing the interchanging process. In general, the Cooper pair operators $C(\lambda)$, $C_{12}(\lambda)$ and $C_{34}(\lambda)$ may take different forms, i.e., their wave functions may be distinct from each other (see below). \\

As we interchange vortex 1 and 2 and hence Majorana zero modes $\bar{\gamma}_1(\lambda)$ and $\bar{\gamma}_2(\lambda)$, all the Cooper pair operators involved in the operator that relates $|00(\lambda)\rangle_{2N}$ and $|11(\lambda)\rangle_{2N}$ can contribute to the Berry phase (more precisely, relative Berry phase between the two projected ground states) as their forms depend on the vortex configuration parameterized by $\lambda$. Again, in the limit of vanishing interchanging area compared to the total system area, all dependence on vortex configuration can be safely ignored and there is no correction to the Berry phase from particle number projection. The interesting question is the possible correction to the Berry phase due to finite trajectory-enclosing area. Ignoring corrections of order $1/N$ as explained below equation (\ref{Berry_Cooper}), the correction depends on whether Cooper pair operators $C(\lambda)$, $C_{12}(\lambda)$ and $C_{34}(\lambda)$ are of the same form. If they were to take the same form, we expect their contribution to the Berry phase to vanish as contribution from $C(\lambda)$ cancels that from the other two Cooper pair operators. Otherwise, we can not rule out finite correction to the Berry phase, which is expected to be scaled as (or some more complicated dependence on $S$ but the magnitude should be bounded by $O(S)$) the dimensionless area of the interchanging region $S$, as implied from analysis in the two-vortex case. \\

Suppose $C(\lambda)$, $C_{12}(\lambda)$ and $C_{34}(\lambda)$ take the same form, it is reasonable to argue that the corresponding BdG ground state $|00(\lambda)\rangle_{\mathrm{BdG}}$ should take the completely paired form: $\mathrm{exp}\{C^\dagger_{00}(\lambda)\}|\mathrm{vac}\rangle$ (up to order $1/N$ correction due to non-commutativity between fermion field operators which are localized in space as Majorana wave functions $u_i(r,\lambda)$ and the Cooper pair operators which are spatially extensive). This is indicated by the same Cooper pair wave function (corresponding to the Cooper pair operators involved in equation (\ref{00_11})) that relates the projected states $|\Psi_n(\lambda)\rangle$ (cf. equation (\ref{BdG_number_space})) in different fermion number sectors with fermion number $2N-2$, $2N$ and $2N+2$, respectively. However, we can prove that in general, the BdG ground states $|00(\lambda)\rangle_{\mathrm{BdG}}$ and $|11(\lambda)\rangle_{\mathrm{BdG}}$ can not be both written in  completely paired form which would otherwise contradict the non-trivial monodromy phase (cf. equation (\ref{delta_alpha})) predicted by the mean-field theory. This can be easily seen as follows. Let's take $|00(\lambda)\rangle_{\mathrm{BdG}}$ and assume it can be written as $\mathrm{exp}\{C^\dagger_{00}(\lambda)\}|\mathrm{vac}\rangle$. After interchanging vortex 1 and 2, the BdG ground state must return back to its initial state, therefore we must have $C^\dagger_{00}(\lambda_f)=\mathrm{exp}\{i\eta_C\}C^\dagger_{00}(\lambda_i)$ (where $\lambda_i$ and $\lambda_f$ label vortex configurations before and after the interchange, respectively). The corresponding BdG ground state after the interchange then takes the form
\begin{eqnarray}
|00(\lambda_f)\rangle_{\mathrm{BdG}}=\sum_n\frac{1}{n!}\mathrm{exp}\{in\eta_C\}C^{\dagger n}_{00}(\lambda_i)|\mathrm{vac}\rangle \label{00_BdG_braid}
\end{eqnarray}

Comparison of the above expression with the initial state written as $|00(\lambda_i)\rangle_{\mathrm{BdG}}=\sum_nC^{\dagger n}_{00}(\lambda_i)/n!|\mathrm{vac}\rangle$ immediately yields the result that $\eta_C$ must vanish (modulo $2\pi$) and hence the monodromy phase for $|00(\lambda)\rangle_{\mathrm{BdG}}$ (An identical argument applies for $|11(\lambda)\rangle_{\mathrm{BdG}}$). As the relative monodromy phase is $\pi/2$ (the same as for the two-vortex case), we conclude that both of the two BdG ground states can not be written in the completely pair form and hence the Cooper pair operators $C(\lambda)$, $C_{12}(\lambda)$ and $C_{34}(\lambda)$ can not all take the same form, leaving the possibility of non-zero contribution to the Berry phase from the Cooper pair operators. \\

In this section, we have shown that as a result of the particle number projection, the braiding statistics of Majorana zero modes may receive finite modification from the Cooper pair contribution. The amount of modification scales as (or bounded by) the area enclosed by trajectories of interchanging Majorana zero modes. This makes braiding Majorana zero modes for quantum computing subject to corrections from the Cooper pair contribution for any realistic system where the area enclosed by interchanging Majorana zero modes is finite (i.e., non-vanishing) in general. Before closing this section, we note an additional issue associated with particle number projection from BdG ground states. In the mean-field theory, the average ground state fermion number is essentially irrelevant to topological properties of Majorana zero modes as long as the system is in the non-trivial topological phase. This makes it questionable for constructing the corresponding particle-number conserving ground states by particle number projection. As shown by a toy model in Appendix \ref{Kitaev}, projected states in different fermion number sectors can have different Berry phases under braiding. This result challenges the conceptual basis of using particle-number non-conserving ground states in the mean-field theory to construct the corresponding particle-number conserving ground states for superfluid systems with fixed fermion number, thereby posing another challenge to the BdG framework for studying topological properties of systems with fixed fermion number. \\


\section{Condensate deformation in the presence of bound Bogoliubov quasiparticles} \label{condensate deform}

Starting in this section, we consider modifications to many-body ground state wave functions described by the mean-field theory which may affect properties of Majorana zero modes. Among various modifications beyond the BdG equations such as Richardson solutions \cite{Richardson} and collective deformation of many-body wave functions resulting from Anderson-Bogoliubov mode (Appendix \ref{sum rules}), we are interested in types of modifications relevant to the topological properties of superfluid ground states. Inspired by the effect of particle number projection considered above, we restrict our considerations to modifications of Cooper pair wave function as it directly affects braiding statistics by contributing to the Berry phase in a particle-number conserving description. Given the current lack of systematic theories beyond the BdG equations, we explore a particular form of many-body wave function ansatz which contains a modified Cooper pair associated with a bound Bogoliubov quasiparticle for a superfluid system where explicit many-body eigenstate wave functions  can be written down in the BdG theory. We show explicitly that the modified ansatz has lower energy than the original ground state wave function in the BdG description. How this modification is related to particle number conservation and its implications for systems with Majorana zero modes will be delayed to section \ref{MZM beyond BdG} for a separate discussion. \\

\begin{figure}[h!]
\begin{center}
\includegraphics[width=0.45\textwidth]{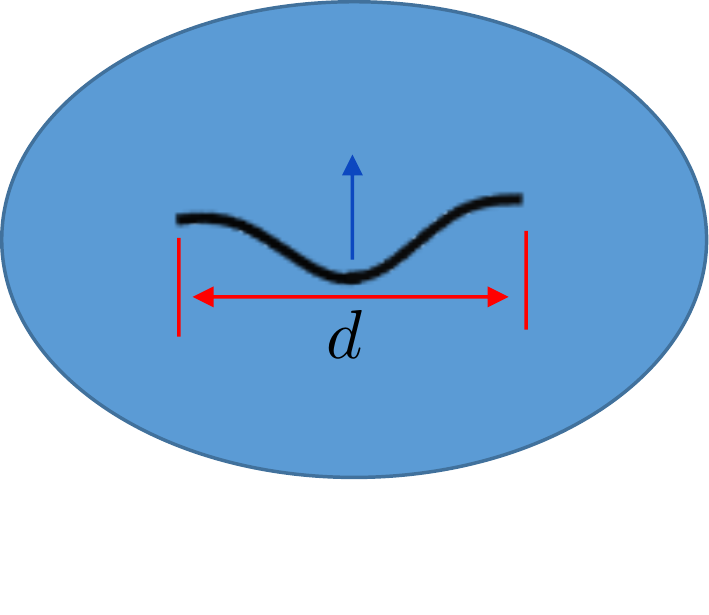}
\end{center}
\vspace{-15pt}
\caption{A s-wave superfluid system with odd number of fermions, bounding a quasiparticle in a weak wide Zeeman trap. The characteristic size of the trap is much larger than superfluid coherent length $d\gg \xi$.} 
\label{fig_bound_pair}
\end{figure}

The system we consider is a superfluid with s-wave pairing containing fixed odd particle number (see also a similar model studied in \cite{Lin_Leggett_1}). A weak external Zeeman field is imposed on a region with linear dimension much larger than the superfluid coherent length, $d\gg \xi$ (see figure \ref{fig_bound_pair}). The strength of the Zeeman field is such that it can bind at least one Bogoliubov quasiparticle mode and at the same time has minimal effect on the form of the superfluid condensate. We consider the system ground state and study whether it is modified from the BdG description by a finite amount due to Cooper pair deformation in the region of the bound Bogoliubov quasiparticle. We consider the possible modification in the thermodynamic limit in which the size of the bound quasiparticle vanishes compared to the system size. \\

In the standard BdG framework, the ground state can be obtained as follows. Owing to the odd particle number parity, the ground state doesn't correspond to the (even number parity) particle-number non-conserving ground state. Instead, it is the first excited eigenstate containing the bound Bogoliubov quasiparticle with the lowest excitation energy. The first excited eigenstate is then projected onto fixed particle number sector to yield the ground state of the odd particle-number system in the BdG approximation. As already discussed in the above section, the particle-number conserving ground state can be related to the particle-number conserving ground state with even particle-number parity by a particle-number conserving Bogoliubov quasiparticle operator. For the system under consideration, the even particle-number parity ground state can be approximated by the homogeneous BCS ground state with vanishing modification for weak Zeeman field strength. The Cooper pair associated with particle-number conserving Bogoliubov quasiparticle has the same form as that constituting the homogeneous BCS ground state written in the completely paired form. This is the assumption made in section \ref{braid}.\\

We now study the possibility of the Cooper pair deformation due to the presence of the bound Bogoliubov quasiparticle. To this end, we propose the modified ground state ansatz to be a linear combination of the ground state obtained in the BdG approximation, as described in section \ref{braid} and in the above paragraph and state modified from the former by deforming the Cooper pair associated with the particle-number conserving Bogoliubov quasiparticle in the region where the quasiparticle is bound. To simplify calculations, we shall satisfy ourselves with the particle-number non-conserving form of the ansatz (which can be projected onto fixed particle number sector for a particle-number conserving form). The BdG many-body wave function with a localized quasiparticle is 
\begin{eqnarray}
\Psi_0=\sum_q \beta_qa^\dagger_{q\uparrow}\prod_{k\neq q}(u_k+v_kb^\dagger_k)|\mathrm{vac}\rangle, \label{Psi_0}
\end{eqnarray}
where the sum over $q$ goes over states close to the Fermi surface, namely those satisfying $\epsilon_q\ll\bigtriangleup$, and where $b^\dagger_k\equiv a^\dagger_{k,\uparrow}a^\dagger_{-k,\downarrow}$. By contrast, the many-body wave function containing one modified Cooper pair is 
\begin{eqnarray}
\Psi_1=\sum_q \beta_qa^\dagger_{q\uparrow}\sum_{q'\neq0}\lambda_{q'}\sum_{k'\neq q\pm q'/2}c_{k'}b^\dagger_{k',q'}\prod_{k\neq q,k'\pm q'/2}(u_k+v_kb^\dagger_k)|\mathrm{vac}\rangle, \label{Psi_1}
\end{eqnarray}
where $q'$ has magnitude no bigger than $d^{-1}\ll\xi^{-1}$ and $b^\dagger_{k',q'}=a^\dagger_{k'+q'/2,\uparrow}a^\dagger_{-k'+q'/2,\downarrow}$. To facilitate quantifying the amount of modification from the BdG ground state, $\Psi_1$ is orthogonal to $\Psi_0$ by construction (the modified Cooper pair in $\Psi_1$ contains pairing of only non-zero center-of-mass momentum, i.e., $q'\neq0$, which is orthogonal to $\Psi_0$ containing only zero center-of-mass momentum pairing). To satisfy the normalization conditions for $\Psi_0$ and $\Psi_1$, $\beta_q$ is of order $1/\sqrt{N_q}$, $\lambda_{q'}$ is of order $1/\sqrt{N_{q'}}$ and $c_{k'}$ is of order $1/\sqrt{N}$, where $N_q$, $N_{q'}$ are number of different $q$ and $q'$ that are summed over, respectively and $N$ is the total average particle number. \\

Now let's consider the matrix element $\langle\Psi_0|H|\Psi_1\rangle$ where the system Hamiltonian $H$ is given by
\begin{eqnarray}
H=\sum_{k,\sigma} \epsilon_ka^\dagger_{k\sigma}a_{k\sigma}-V_0\sum_{k,k',q}b^\dagger_{k',q}b_{k,q}, \label{H}
\end{eqnarray}
where we have omitted the Zeeman term for simplicity. A straightforward calculation yields the following estimate
\begin{eqnarray}
\langle \Psi_0|H|\Psi_1\rangle&=&V_0\sum_{q,q',k'}\beta_{q}\beta_{q+q'}\lambda_{q'}c_{k'}(v_qu_{k'+q'/2}u_{k'-q'/2}-u_qv_{k'+q'/2}u_{k'-q'/2}) \nonumber \\
&\sim&V_0\frac{1}{N_q}\frac{1}{\sqrt{N_{q'}}}\frac{1}{\sqrt{N}}N_qN_{q'}N\frac{\bigtriangleup}{\epsilon_F} \nonumber \\
&\sim&V_0\sqrt{NN_{q'}}\frac{\bigtriangleup}{\epsilon_F} \nonumber \\
&\sim&\bigtriangleup\sqrt{N_{q'}/N}. \label{off_diag_ME}
\end{eqnarray}

We see that the matrix element is independent of the system size. In fact, the actual value is larger by noticing that $c_{k'}$ in (\ref{Psi_1}) is constrained in a thin shell around the Fermi surface, since only single particle sates near the Fermi surface can be scattered by the BCS potential. The coupling between $\Psi_0$ and $\Psi_1$ is due to scattering of pairs of particles in $\Psi_0$ by non-zero center-of-mass momentum scattering terms $-V_0\sum_{k,k',{q\neq0}}b^\dagger_{k',q}b_{k,q}$. \\

We expect that the diagonal energy difference between $\langle\Psi_0|H|\Psi_0\rangle$ and $\langle\Psi_1|H|\Psi_1\rangle$ to be independent of the system size since $\Psi_1$ is different from $\Psi_0$ only by a localized Cooper pair. Since both this energy difference and the off-diagonal energy $\langle \Psi_0|H|\Psi_1\rangle$ are independent of the system size, the variational ground state is superposition of the two states $\Psi_0$ and $\Psi_1$ whose weights are independent of the system size. Hence we reach the conclusion that modification of the Cooper pair associated with the bound quasiparticle is finite and may be regarded as partially bound to the same region in the thermodynamic limit. Note that this partial localization of the Cooper pair is a many-body effect since in a single-particle in a potential well Schrodinger problem, the eigenstate of the particle is either bound to the well or extensive, but never in between. \\

Intuitively, the partially bound Cooper pair may be understood as follows. Imagine we start with the mean-field many-body eigenstate with one localized quasiparticle and consider its evolution in time. When a Cooper pair is near the region of localization, it gets deformed due to Cooper pair-quasiparticle interaction. The percentage of time spent by a Cooper pair near the localization region is vanishing in the thermodynamic limit. However, all Cooper pairs get the same modification, so the effect from all Cooper pairs adds up to compensate the small deformation of each Cooper pair. From this perspective, the more physical ansatz beyond the BdG construction should modify all Cooper pairs in the condensate, not just one pair associated with the quasiparticle hole. \\

We have shown that in general, superfluid condensate can have finite local deformation due to a localized quasiparticle. A most interesting question relevant to superfluid topological properties is whether such a deformation can affect any local physical observable. We shall focus on the simplest observable, the particle number density which is the subject of the next section. \\

\section{Particle localization in Zeeman-Josephson model} \label{particle localization}

Knowing that the superfluid condensate is deformed doesn't directly tell us what happens to local particle number density. In fact, it's impractical to calculate the particle number density directly from the many-body eigenstate (even if we know it) in the presence of a localized quasiparticle, which would take very complicated form. Furthermore, even if we can evaluate it numerically, it wouldn't tell us much about the underlying physics. So we shall look for physics related to particle number conservation, which is the key ingredient not accounted for in the BdG framework. \\

One type of simple model suitable for our consideration is a superconducting charge qubit \cite{Clarke} involving Josephson junction. We consider a system made of two superconducting grains joined by a Josephson junction. The total particle number of the system is conserved. The effective Hamiltonian can be written as
\begin{eqnarray}
H=E_C(n-n_g)^2-E_J\mathrm{cos}\phi. \label{Josephson}
\end{eqnarray}
The first term corresponds to charging energy which depends on particle number change $n$ between two superconducting grains due to Cooper pair transfer between the two sides (we have rescaled $E_C$ to make $n$ particle number change between two superconducting grains, note that in the original definition, $n$ refers to Cooper pair transfer between two superconducting grains, see more detailed explanation below). The second term is responsible for Josephson tunneling energy, $\phi$ is relative phase difference between the two superconducting grains. $n_g$ is offset induced by gate voltage in the original transmon model and it doesn't need to take integer values. This simple Hamiltonian has been solved exactly \cite{Koch}. \\

\begin{figure}[h!]
\begin{center}
\includegraphics[width=0.3\textwidth]{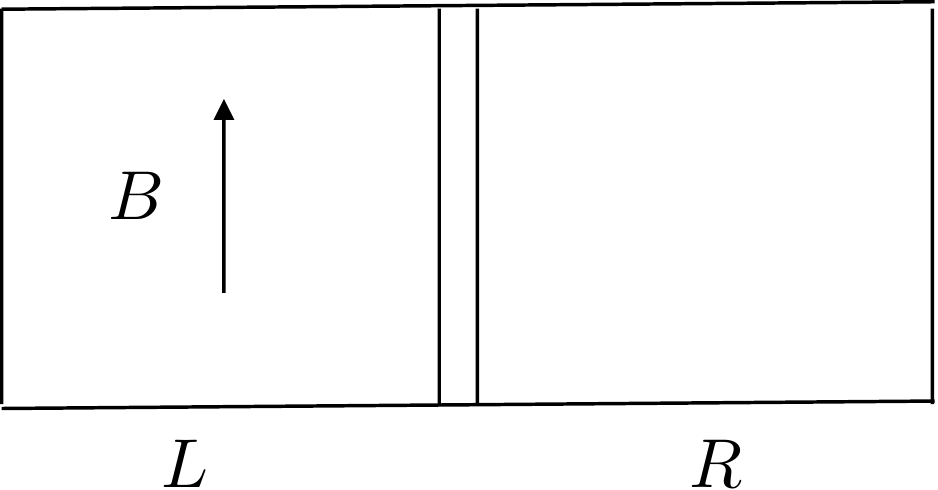}
\end{center}
\caption{Josephson-Zeeman model. We consider two superconducting grains $L$ and $R$ connected by a insulating junction. The total number of particles is odd. A quasiparticle with spin up is localized in the left superconducting grain due to a uniform Zeeman field imposed on the left grain. }
\label{fig_Josephson-Zeeman}
\end{figure}

For our purpose,  we would like to apply the above Hamiltonian to describing the same system, but with total odd number of particles and a uniform weak Zeeman field acting on the left grain (see figure \ref{fig_Josephson-Zeeman}). In other words, we want to study the quasiparticle localized by a Zeeman field in the context of a Josephson junction. The effective Hamiltonian (\ref{Josephson}) is a minimal Hamiltonian describing the competition between the first term which tends to fix the particle number on each grain separately and the second term, which favors particle number fluctuation. We believe this model is relevant to capture the essential physics of particle number localization. \\

To adapt the Hamiltonian (\ref{Josephson}) to our Zeeman-Josephson system, we need to make some assumptions. We need to assume that the unpaired particle contributes to the charging energy equally as that of each paired particle so that it doesn't change $E_C$. We further assume that the Zeeman field is such that the unpaired particle is only residing on the left side, namely, we ignore single particle tunneling process across the Josephson junction. Under these assumptions, the left side grain is in a mixture of odd number of particles whereas the right side is in a mixture of even number of particles. In the absence of the unpaired particle (when the system has even number of particles), we set $n_g=0$. In the presence of the unpaired particle, we assume its effect is to change $n_g$ to 1 for the following reason. In the infinite $E_J$ limit, we expect no extra particle localized on the left grain due to the localized quasiparticle, but only a net spin localized on the left. In other words, the average particle number is increased by one half on both sides due to one added particle to the even particle-number ground state. So the expectation value of $n$ in the ground state should be equal to 1 ($\langle n\rangle=1$), namely on average, a quarter of Cooper pair (that is half a particle) is transferred from the left to the right to compensate one unpaired particle added to the left. (Recall that $n$ refers to particle number change between the two sides. So for example, two particles transferred from left to right are equivalent to a four-particle change, $n=4$.) Therefore $n_g=1$ according to Hamiltonian (\ref{Josephson}) since in the infinite $E_J$ limit, $\langle n\rangle=n_g$. \\

Having justified using an effective Hamiltonian (\ref{Josephson}) with $n_g=1$ to calculate the ground state of our Zeeman-Josephson system with odd number of particles, we can make use of available analytic results \cite{Koch} to evaluate the average particle number change $\langle n\rangle$ for finite but large $E_J/E_C$. By the Feynman-Hellman theorem, $\langle n\rangle$ in the ground state can be related to ground state energy derivative
\begin{eqnarray}
\langle n\rangle&=&\frac{1}{2E_c}\langle\frac{\partial H}{\partial n}\rangle+n_g \nonumber \\
&=&-\frac{1}{2E_c}\langle\frac{\partial H}{\partial n_g}\rangle+n_g \nonumber \\
&=&-\frac{1}{2E_c}\frac{\partial E_0}{\partial n_g}+n_g,  \label{n_FH}
\end{eqnarray}
where $E_0$ is the ground state energy. \\

We can now apply the analytical solutions found by Koch et al. \cite{Koch} to evaluate (\ref{n_FH}). Noticing the their definition of the energy scale $E_C$ and ours, the derivative $\partial E_0/\partial n_g$ is actually taken at $n_g=1/4$ in their definition. We get the following estimate for $\langle n\rangle$
\begin{eqnarray}
\langle n\rangle = 1-f(E_J/E_C)\mathrm{exp}(-\sqrt{2E_J/E_C}), \label{n_est}
\end{eqnarray}
where $f(E_J/E_C)>0$ is a power law function of $E_J/E_C$ whose specific form is not very important here. \\

We see from equation (\ref{n_est}) that the particle number transfer by Cooper pairs from left to right is smaller than one half by amount which is exponentially small. The upshot is that when finite charging energy $E_C$ is taken into account, there is net particle localization on the left due to the localized quasiparticle. If we were to do the calculation in the BdG framework, the extra particle (relative to system with even particle number) would distribute evenly on the two sides. The particle and hole component of the localized quasiparticle has the same weight and their contribution to the total particle number on the left cancels. Including the Cooper pair to enforce fixed particle number yields a uniform distribution for one extra particle since the Cooper pair is what constituting the even particle-number parity ground state which approaches homogeneous BCS ground state for a weak Zeeman field in the BdG framework. Therefore, the BdG framework corresponds to vanishing charging energy in our model. By contrast, we have shown that particle number conservation can lead to net particle localization due to charging energy. \\

It is still nontrivial to generalize the above result to actual situations of a localized quasiparticle in superfluids. But intuitively, we can think of the region of the localized quasiparticle as one superconducting grain and the rest as the other grain. Finite compressibility induces a charging energy cost in particle number fluctuation in each region whereas Josephson energy favors particle number fluctuation. In the thermodynamic limit, the ratio of Josephson energy to charging energy stays finite and hence we expect particle localization to scale as inverse power law of total system size. The power law decay is determined by the ratio of total fermion number in the localized region to that in the whole system. This ratio determines $n_g$ at which the derivative $\partial E_0/\partial n_g$ (cf. equation ({\ref{n_FH})) is evaluated. $n_g$ is proportional to the ratio and $\partial E_0/\partial n_g$ evaluated at $n_g$ decays in power law as the ratio goes to zero (in the above analysis leading to equation (\ref{n_est}), we have implicitly assumed equal size for the two superconducting grains so that the ratio is 1 there). Of course, there are still very critical differences between a localized quasiparticle in superfluids and the Zeeman-Josephson model. In actual situations of interest, there's no real Josephson junction separating the two regions and no constraints on particle number parity in either region. It is much more difficult to evaluate the effect of particle localization in more physically relevant systems and we leave it to future study. The implication of the particle localization due to a bound quasiparticle to detectability of Majorana zero modes will be discussed in the next section. \\

\section{Effect of particle number conservation on Majorana zero modes}  \label{MZM beyond BdG}

In this section, we discuss possible effect of particle number conservation on Majorana zero modes. We have shown in the above two sections by simple physical systems that a bound quasiparticle can induce local condensate deformation and possible particle localization that are not accounted for in the standard BdG framework. Now the most relevant questions involve whether similar effects exist for superfluid ground states in the presence of Majorana zero modes and how these effects may change the topological properties of Majorana zero modes as predicted in the BdG framework. \\

Let's first consider the following question: whether Cooper pairs in degenerate superfluid ground states spanned by Majorana zero modes receive relative deformation as a consequence of particle number conservation. We are interested in the relative deformation of degenerate ground states as the braiding statistics depends on the relative Berry phase of braiding Majorana zero modes (cf. equation (\ref{total phase})). To address the possible many-body eigenstate deformation beyond the BdG approximation due to particle number conservation, it is worth first considering particle number conservation within the BdG framework (see Appendix \ref{continuity_BdG} for more details). In general, the continuity equation is not satisfied for an eigenstate of BdG Hamiltonian (it is of course identically satisfied if the exact Hamiltonian is used in evaluation). Instead, it is satisfied only for BdG ground states evaluated at expectation level or for thermal average at non-zero temperatures \cite{Furusaki, Bagwell,Sols}. In particular, it is violated for a general excited eigenstate of the BdG Hamiltonian. However, there is a straightforward way to recover the continuity condition for excited eigenstates within the BdG framework. Namely, the gap equation needs to be modified to take proper account of the occupation of quasiparticles corresponding to the eigenstate in consideration. In the modified BdG framework, the gap, the excitation spectrum and the BdG solutions become all dependent on which eigenstate we consider. For a typical low-lying excited eigenstate, the modification of the gap is only of order $1/N$ (with $N$ total particle number) and hence in most situations, such a modification is negligible for most purposes such as calculating energy spectrum. However, for physical quantities of our interest such as the Berry phase which are directly determined by many-body eigenstates, we may not be able to neglect the modification to the gap as it can result in $O(1)$ modification to the corresponding many-body eigenstate. Specifically, for the system considered in section \ref{condensate deform}, the bound quasiparticle can induce an $O(1)$ modification to the local gap, which in turn modifies the corresponding many-body ground state with odd particle number parity (which is the first excited eigenstate in particle-number non-conserving BdG framework) by a finite amount. Now comes the subtlety concerning Majorana zero modes. Due to the topological protection of Majorana zero modes within the BdG framework, degenerate ground states spanned by Majorana zero modes receive no relative modification from the modified BdG framework since the local gap is independent of whether Majorana zero modes are occupied (or equivalently which degenerate ground state we are considering). Therefore, it is worth seeking evidence for modifications to the Cooper pair that are distinct from the modification given by the modified BdG framework. In this respect, we note that the modification of the ground state in the modified BdG framework is similar in nature to the modification we proposed in the ansatz (\ref{Psi_1}). Furthermore, it is not obvious whether the energy associated with the ansatz (\ref{Psi_1}) is lower than that corresponding to the modified BdG description. Thus we are unable to determine at this stage whether the modification we considered in section \ref{condensate deform} truly goes beyond the BdG framework including the modified BdG framework. Therefore, the question concerning possible deformation of Cooper pair involving Majorana zero modes is still left open. (However, we do find evidence supporting a modification to the many-body ground states beyond the description of the modified BdG framework which is relevant to Berry phase \cite{Lin_Leggett_1}. See also more details on this in the next section.) \\

A similar difficulty applies to the problem of particle localization due to Majorana zero modes. For the Zeeman-Josephson system considered in the above section, we are unable to immediately exclude the possibility that a modified BdG description may yield similar particle localization. However, we may argue in favor of particle localization due to Majorana zero modes. The physics of particle localization in the Zeeman-Josephson model is due to the interplay between odd particle number parity in the localized region and particle number conservation which gives rise to the charging energy. Since the local regions where Majorana zero modes reside may have different particle number parity expectation values for different degenerate ground states, it is possible that relative particle localization may occur in degenerate ground states (given the caveat concerning the distinction between a bound quasiparticle in general superfluids and the Zeeman-Josephson system). \\

We now comment on how the topological properties of Majorana zero modes may be affected if Cooper pair deformation and particle localization were to occur in degenerate ground states. The Cooper pair deformation may modify the Berry phase significantly and hence affect braiding statistics of Majorana zero modes in non-trivial ways, which may change or even spoil Majorana-based quantum computing. There is also another aspect about the Cooper pair deformation which is more speculative: Local condensate deformation could potentially change the degree of freedom associated with Majorana zero modes as long-range entanglement in the wave functions of degenerate ground states may be destroyed. This can happen since local condensate deformation tends to make many-body states acted on by Majorana zero modes at different locations to become orthogonal. (This can be illustrated by the following simple example. Consider the doubly degenerate ground states $|0\rangle$ and $|1\rangle$ related by $|1\rangle=1/2(\gamma_1+i\gamma_2)|0\rangle$ where $\gamma_1$ and $\gamma_2$ are two Majorana zero modes located at separate positions. The long-range entanglement in the degenerate ground states can be seen from relations $\gamma_1|0\rangle=i\gamma_2|0\rangle$ and $\gamma_1|1\rangle=-i\gamma_2|1\rangle$ since $\gamma_1$ and $\gamma_2$ are local operators acting at separate locations. If there's local condensate deformation at the regions where they act, we may get $\gamma_1|0\rangle$ to be orthogonal to $\gamma_2|0\rangle$ and similarly for ground state $|1\rangle$.) If this turns out to be the case, we no longer have topologically ordered many-body eigenstates in the first place. On the other hand, the particle localization may result in equally, if not more, serious consequences to topological quantum computing with Majorana zero modes. Particle localization implies local particle number density distinction for degenerate ground states, thus removing topological protection. Based on discussions in the last paragraph of the above section, we conjecture that particle localization decays as power law of separation between Majorana zero modes. Therefore, any general local perturbation splits the degenerate ground state energy levels to power law decay, instead of exponential decay. Particle localization also has interesting implication to Majorana zero modes in hybrid systems, most notably proximity-induced 1D topological superconducting wires with strong spin-orbit coupling under external magnetic field \cite{Lutchyn}. In such a system, we may regard proximity-induced superconducting wire(s) as superconducting island(s) and the superconducting substrate as the other island. In such an approximation, the hybrid system may be viewed as a Josephson junction between the wire(s) and the superconducting bath in the strong coupling limit. Provided such a approximation is legitimate, the particle localization in the wire(s) will scale as inverse power law of the size of the superconducting bath (relative to that of the wire(s)), multiplied by an exponential factor which decays in wire length(s). The length scale(s) will set by wire charging energy and coupling energy between the wire(s) and the superconducting bath. The dependence of particle localization and hence energy splitting between degenerate ground states due to particle number conservation differs from standard predictions in the BdG framework and may be tested experimentally.

\section{Summary} \label{summary}

In this paper, we have made a first attempt towards constructing a particle-number conserving theory of Majorana zero modes in p+ip superfluids. We emphasize that particle number conservation is necessary for deriving physically sensible results on the braiding statistics of Majorana zero modes in condensed matter  systems in which fermion number is always conserved. We have managed (with modest success) to show that particle number conservation could have important effects due to contribution from Cooper pairs in the condensate. \\

We started within the BdG framework and investigated whether particle-number conserving many-body states as projected from particle-number non-conserving many-body eigenstates in the BdG framework inherit the same topological properties for Majorana zero modes. We found that while topological properties remain unchanged in the thermodynamic limit, they are subject to modifications for realistic systems of finite size. We demonstrated the possibility of modifications to the braiding statistics of Majorana zero modes in vortices of p+ip superfluids due to the Cooper pair contribution.\\

We then considered more subtle effect as consequences of particle number conservation by exploring modifications to many-body wave functions beyond BdG description. We focused on the particular type of modifications manifest in Cooper pair/condensate deformations near bound quasiparticles, motivated by the observation that Cooper pairs can contribute directly to Berry phase. We showed that local modifications to the Cooper pairs due to a bound quasiparticle can yield finite modifications to many-body eigenstates of BdG Hamiltonian. Further manifestation of particle number conservation is seen in local physical observable modification near a bound quasiparticle such as induced particle localization due to finite charging energy. \\

The generalization of these modifications to Majorana zero modes is highly nontrivial, and we need to have a much better understanding of the underlying many-body ground states beyond BdG approximation. At the time of writing, we are unable to distinguish the modification to Cooper pairs we considered from that due to a modified BdG framework, which has no effect on Majorana zero modes.  However, it's suggestive of such modifications beyond modified BdG framework, as evidenced in the 1D annulus model in which the Berry phase resulting from transport of a bound quasiparticle is studied by the present authors (see the description of the model in Introduction). We found that modification to the ground state wave function involves entanglement between the bound quasiparticle and the superfluid condensate, which can not be accounted for even in the modified BdG framework. Such a beyond-BdG modification to the ground state is necessary for obtaining a physically sensible Berry phase compatible with f-sum rule (which is closely related to particle number conservation). As for the particle localization, we have provided arguments in favor of its existence since relevant physics depends on local particle number parity and should be applicable to Majorana zero modes as well. Should such modifications to occur to Majorana zero modes, they may change the topological properties of Majorana zero modes significantly and have profound consequences on topological quantum computing with Majorana zero modes. \\

Throughout our analysis, we see that the physics of Majorana zero modes in superfluids deserves much more sophisticated studies beyond simple BdG framework which breaks particle number conservation. In this report, we have just launched the first attempt seeking a particle-number conserving theory of Majorana zero modes and we are looking forward to more such efforts to either verify or drastically modify the current physical picture of Majorana zero modes in superfluids.


 \appendix
 \appendixpage
\begin{appendices}
\section{BCS mean-field theory and sum rules} \label{sum rules}

In this appendix, we illustrate some unphysical consequences of the BCS mean-field theory as a result of its violation of particle number conservation. The BCS mean-field theory and the corresponding reduced BCS Hamiltonian violates gauge invariance \cite{Anderson} and this problem came under intensive debate in the early days of BCS theory. It was realized that the collective modes of the condensate need to be included to restore gauge invariance and various sum rules such as the f-sum rule. In the case of a translational invariant system, pair interactions with non-zero center-of-mass momenta need to be included, as shown by Anderson, to enforce gauge invariance, and we get low energy collective excitations such as the Anderson-Bogoliubov (AB) mode \cite{Anderson_1} for neutral superconductors. Here, we show that the homogeneous BCS ground state wave function violates a sum rule and explore how one might modify it, followed by an example in  Appendix \ref{example} in which a gauge invariant calculation taking proper account of the condensate dynamics is essential. Finally in Appendix \ref{Galilean}, we provide another consequence of violating particle number conservation that results in violation of Galilean invariance. \\

The well-known homogenous BCS ground state wave function is given by
\begin{eqnarray}
|\mathrm{GS}\rangle_{\mathrm{BCS}}=\prod_k(u_k+v_ka^\dagger_ka^\dagger_{-k})|\mathrm{vac}\rangle, \label{GS_BCS}
\end{eqnarray}
where $u^2_k=1/2(1+\epsilon_k/E_k)$ and $v^2_k=1/2(1-\epsilon_k/E_k)$, $\epsilon_k$ is the single particle kinetic energy defined relative to the Fermi energy, $E_k=\sqrt{\epsilon^2_k+\bigtriangleup^2}$ is the quasiparticle energy spectrum. For simplicity, we have omitted spin indices. Let's evaluate the long-wavelength density fluctuations $S_q\equiv\langle\rho_q\rho_{-q}\rangle$ in the BCS ground state: This can be calculated by expanding $\rho_q$ and $\rho_{-q}$ in terms of Bogoliubov quasiparticles $\alpha^\dagger_{k\sigma}$ with energy $E_{k\sigma}=\sqrt{\epsilon_k^2+|\bigtriangleup|^2}$ and making use of $\langle\alpha_{k\sigma}\alpha^\dagger_{k'\sigma'}\rangle=\delta_{kk'}\delta_{\sigma\sigma'}$, $\langle\alpha_{k\sigma}\alpha_{k'\sigma'}\rangle=\langle\alpha^\dagger_{k\sigma}\alpha^\dagger_{k'\sigma'}\rangle=0$.
\begin{eqnarray}
\lim_{q/k_F\rightarrow0}\langle\rho_q\rho_{-q}\rangle\simeq 2\pi\bigtriangleup N(0), \label{density_fluc_BCS}
\end{eqnarray}
where $N(0)$ is the density of states at Fermi surface and $\bigtriangleup$ is the superconducting gap. 
Expression (\ref{density_fluc_BCS}) approaches a constant in the long wavelength limit; this is noted in \cite{Anderson}, but surprisingly it is not remarked there that this behavior violates the sum rules (f-sum rule and compressibility sum rule). We can see this as follows: the upper bound of the long wavelength density fluctuation can be found from a Cauchy-Schwartz inequality combined with f-sum rule and compressibility sum rule as
\begin{eqnarray}
\lim_{q/k_F\rightarrow0}\langle\rho_q\rho_{-q}\rangle<\frac{Nq}{mc}, \label{dens_fluct_upper}
\end{eqnarray}
where $N$ is the total particle number and $c$ is the speed of the low-energy hydrodynamic mode which is of the order of the Fermi velocity. So the long-wavelength density  fluctuations should vanish as $q$ approaches zero. \\

We may modify the BCS ground state wave function by including the zero-point fluctuations of the AB modes by the following ansatz (which is actually rather routine in first-principles approaches to neutral Fermi systems, see e.g., \cite{Feenberg})
\begin{eqnarray}
|\mathrm{GS}\rangle_{\mathrm{mod}}= \mathrm{exp}(-\sum_q\lambda_q\rho_q\rho_{-q})|\mathrm{GS}\rangle_{\mathrm{BCS}}, \label{GS_mod}
\end{eqnarray}
where $\lambda_q=mc/(Nq)$.\\

In the above ansatz, long-wavelength density fluctuations are damped by $\lambda_q$ as $q$ approaches zero. \\

\subsection{An Example of Failure of Application of the BdG equations - NMR signature of $^3$He-B Surface} \label{example}

Here we present an example where a gauge invariant calculation taking into account superfluid condensate dynamics is essential in obtaining correct physics. We consider NMR longitudinal absorption in a $^3$He-B film. It is well known that for bulk $^3$He-B, the longitudinal absorption is solely due to nuclear spin dipole interaction which is the only source in the Hamiltonian breaking rotation symmetry of spin relative to orbital coordinates, and so resonance peak is completely determined by dipolar energy, whose dominant contribution comes from the superfluid condensate spins due to their collective behavior (see e.g.,\cite{Leggett_1} and \cite{Leggett_2}, for a review, see \cite{Leggett_3}). Including the surface shouldn't qualitatively change the nature of the NMR response. However, if we implement mean-field calculation as done in \cite{Silaev}, we will get a qualitatively wrong signal in which absorption starts from the surface BdG quasiparticle energy gap which is dependent on the Larmour frequency. In particular, in the limit of vanishing external magnetic field, the absorption as calculated in \cite{Silaev} starts from zero frequency, signaling Majorana zero modes localized at the surfaces. In this example, we see that if the condensate has interesting internal structure which is spin structure in this case, we need to pay attention to the condensate dynamics and shouldn't without argument take the condensate as  c-number background. Technically, in this example, a mean-field calculation based on the BdG equations breaks the conservation laws. A gauge invariant calculation (for gauge invariant schemes see e.g. \cite{Baym}) has been done by Taylor et al. \cite{Taylor} and they confirmed that the qualitative feature of NMR absorption is unchanged from that of the bulk. See figure \ref{fig_NMR} for an illustration of NMR longitudinal absorption of a $^3$He-B film. \\

\begin{figure}[h!]
\begin{center}
\includegraphics[width=0.7\textwidth]{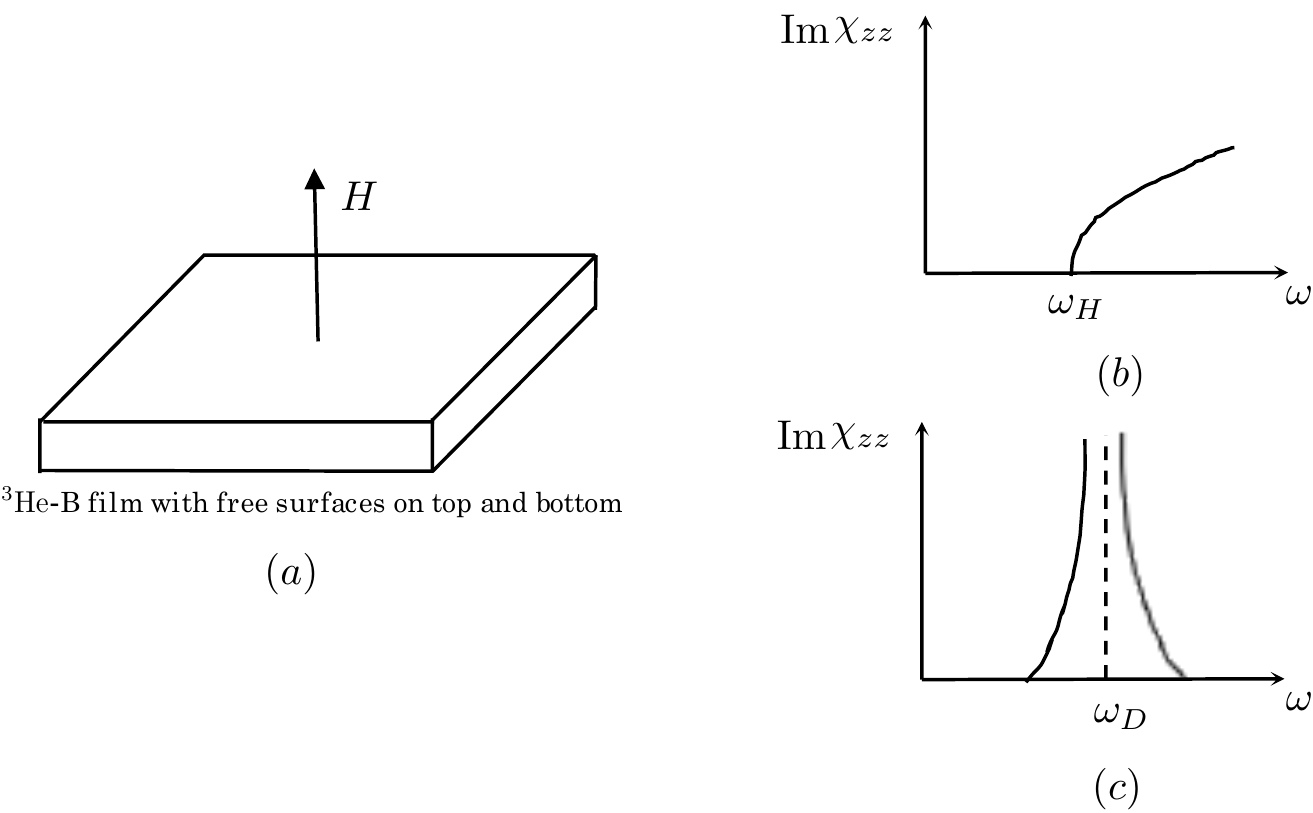}
\end{center}
\vspace{-15pt}
\caption{NMR longitudinal absorption of a $^3$He-B film. (a) system setup, magnetic field $H$ is uniform pointing in z direction (b) prediction based on mean-field BdG solutions: the absorption occurs at Larmour frequency $\omega_H$ (c) result based on gauge invariant calculation by Taylor et al.: the absorption signal agrees qualitatively with that of the bulk with resonance frequency determined by nuclear spin dipole energy $\omega_D$ and broadened by Majorana zero modes at the film surface, sketch is adapted from figure 3 of \cite{Taylor}.}
\label{fig_NMR}
\end{figure}

\subsection{Galilean Invariance} \label{Galilean}

The lack of particle number conservation in the BCS theory also results in violation of Galilean invariance: we illustrate this by comparing the momentum of a BdG quasiparticle in two reference frames, one boosted from the other by a finite velocity. The BdG quasiparticle is created in a BCS s-wave uniform superfluid 
\begin{eqnarray}
\alpha^\dagger_k=u_ka^\dagger_{k\uparrow}+v_ka_{-k\downarrow}, \label{BdG_k}
\end{eqnarray}
with the corresponding BCS ground state wave function taking the standard BCS form $|\mathrm{GS}\rangle=\prod_k(u_k-v^*_ka^\dagger_{k\uparrow}a^\dagger_{-k\downarrow})|\mathrm{vac}\rangle$ in the lab frame.
We first calculate in the lab frame the momentum of the quasiparticle described by (\ref{BdG_k}) by comparing the total momentum of the superfluid with and without the quasiparticle excitation
\begin{eqnarray}
p_{\mathrm{lab}}=\langle\alpha_k P \alpha^\dagger_k\rangle-\langle P\rangle=\langle\alpha_k,[P,\alpha^\dagger_k]\rangle. \label{p_lab}
\end{eqnarray}
The commutator of total momentum operator $P$ with $\alpha^\dagger_k$ is evaluated to be $k\alpha^\dagger_k$. Inserting it back to the above equation, we get $p_{\mathrm{lab}}=k$. This is what we expect. \\

Now we boost the superfluid by switching to a moving frame in which the ground state wave function becomes $|\mathrm{GS}'\rangle=\prod_k(u_k-v^*_ka^\dagger_{k+K/2\uparrow}a^\dagger_{-k+K/2\downarrow})|\mathrm{vac}\rangle$ with total momentum of each Cooper pair equal to $K$. The corresponding quasiparticle creation operator in the boosted superfluid is given by
\begin{eqnarray}
\tilde{\alpha}^\dagger=u_ka^\dagger_{k+K/2\uparrow}+v_ka_{-k+K/2\downarrow}. \label{BdG_k_K}
\end{eqnarray}

The quasiparticle momentum in the boosted frame is evaluated in a similar way to be
\begin{eqnarray}
p_{\mathrm{boost}}=k+\frac{K}{2}(|u_k|^2-|v_k|^2). \label{p_boost}
\end{eqnarray}

This result violates the Galilean invariance according to which we expect the quasiparticle momentum in the boosted frame to be $k+K/2$. This is simply due to particle-number non-conserving form of the many-body wave functions. As we compare the expectation value of total particle number of many-body eigenstates with and without the quasiparticle excitation, the particle number is not increased by one, but by $|u_k|^2-|v_k|^2$. Therefore in the boosted frame, the momentum change due to the quasiparticle is not equal to $k+K/2$. The resolution for fixing this problem is either to adjust coefficient of $u_k$ and $v_k$ for all k when the quasiparticle is added to the superfluid ground state to ensure average total particle number is increased by one (recall that the average total particle number is determined by $\sum_k|v_k|^2$, by adjusting $v_k$ (and also $u_k$ for normalization condition) we can ensure the desired average total particle number) or simply to associate a Cooper pair creation operator to the hole part of the quasiparticle. If we associate a Cooper pair to the hole part of the quasiparticle, the boosted particle momentum satisfies Galilean invariance
\begin{eqnarray}
p'_{\mathrm{boost}}&=&k+\frac{K}{2}(|u_k|^2-|v_k|^2)+K|v_k|^2  \nonumber \\
&=&k+\frac{K}{2}(|u_k|^2+|v_k|^2)=k+\frac{K}{2},\label{p'_boost}
\end{eqnarray}
where the last term on rhs of first equality above is contribution from the Cooper pair to the momentum. This result satisfies Galilean invariance. \\

This simple example illustrates possible violation of fundamental physical principles due to particle-number non-conserving in the standard mean-field approach. In this particular example, we see that extra care needs to be taken when the superfluid is moving. \\

\section{Braiding Majorana fermions using 1D Kitaev Network with Conserved Particle Number} \label{Kitaev}

In this appendix, we lay out the details of calculating the braiding statistics of Majorana zero modes in 1d Kitaev-wire networks with conserved particle number. Since we consider an effectively quasi-1d system (three-dimension physically to support superconducting long-range order) with fixed particle number (i.e., it doesn't exchange particles with environment) and the original Kitaev-wire \cite{Kitaev_wire} doesn't conserve particle number, we use the Kitaev-wire Hamiltonian (i.e., mean-field BCS Hamiltonian) to first calculate particle-number non-conserving many-body eigenstates and then project onto a fixed particle-number sector to obtain particle-number conserving states at the mean-field level. The motivation for doing this is to check whether states with fixed particle number evolve under interchanging Majorana zero modes in the same way as particle-number non-conserving states do. The monodromy phase will not be affected by fixing particle number since states with fixed particle number are obtained by projecting particle-number non-conserving states. However, the Berry phase may differ for the two cases as the Berry phase for a particle-number non-conserving state is the average Berry phase of states each with fixed particle number. This average is not necessarily equal to its constituents unless there is an equal contribution from every state with fixed particle number (cf. Section \ref{braid}). \\

In order to move Majorana zero modes, we impose some external potentials on the wires so that both chemical potential and particle-particle interactions can be tuned. To keep the calculation physical, there are two criteria to meet during the move.  \\

1. The self-consistency condition or gap equation needs to be satisfied. This requirement may complicate calculations if we had to iterate mean-field calculations to achieve self-consistency. Furthermore, we require physically reasonable particle-particle interactions (in our calculations, a short-range interaction). \\

2. The average particle number calculated by the mean-field approach needs to be fixed during the braiding. Since we are considering particle-number conserving wave functions, the average particle number needs to remain constant before projection. \\

There is an issue related to the second criterion also discussed in the main text. The second criterion is to ensure that particle-number conserving wave functions are well approximated by the particle-number non-conserving wave functions. In large systems with macroscopic particle number, this condition may be loosened to allow certain amount of variation in average particle number. It is unclear to what extent this can be done without affecting the braiding statistics.  \\

Furthermore, there's another issue, namely orthogonality. As particle-number conserving wave functions are obtained by projection from particle number non-conserving wave functions which form an orthonormal basis, after projection the wave functions may be slightly non-orthogonal. To calculate the braiding statistics self-consistently, we need to use an orthonormal basis. As the system size becomes large, the degree of non-orthogonality should become small. However, it is not totally clear whether the non-orthogonality decreases fast enough to ensure consistency. \\

With these caveats, let's proceed to examine braiding Majorana zero modes in fixed particle number systems. We'll first check in \ref{Criteria Check} whether the two criteria can be realized in a simple system. Before embarking on braiding Majorana fermions, we'll briefly review in \ref{Non-Abelian} non-Abelian braiding with an emphasis on gauge invariance which will be useful for simplifying calculation in exchanging two Majorana zero modes. Finally, in \ref{Double} and \ref{Single}, we discuss double and single interchange braiding scheme respectively. \\

\subsection{Criteria Check} \label{Criteria Check}
Fortunately, the two criteria can be satisfied in a relatively straightforward way. Consider a single wire with two open ends described by the following mean-field particle-number non-conserving Hamiltonian (i.e., Kitaev-wire Hamiltonian)
\begin{eqnarray}
H=\sum_{i=1}^{N-1}(a^\dagger_i+a_i)(a^\dagger_{i+1}-a_{i+1}). \label{H_single_wire}
\end{eqnarray}
Defining Majorana fermions as 
\begin{eqnarray}
\gamma_A&=&\frac{a^\dagger-a}{i}, \nonumber \\
\gamma_B&=&a^\dagger+a \label{gamma},
\end{eqnarray}
equation (\ref{H_single_wire}) can be written as
\begin{eqnarray}
H=\sum_{i=1}^{N-1}i\gamma_{i,B}\gamma_{i+1,A}. \label{H_single_wire_MF}
\end{eqnarray}
There are two Majorana zero modes at the ends $\gamma_{1,A}$ and $\gamma_{N,B}$. It can be easily checked that the order parameter in the ground state is 
\begin{eqnarray}
\langle a_{i+1}a_i\rangle=-1/4, \label{OP_single_wire_0}
\end{eqnarray}
for any $N>2$, where $i=1,\cdots,N-1$. 
To satisfy the gap equation, we only need to require the nearest-neighbor interaction to be
\begin{eqnarray}
V_{i,i+1}=-4. \label{V_single_wire_0}
\end{eqnarray}
So a physically realistic nearest-neighbor interaction suffices to satisfy the self-consistency condition. The full particle-number conserving Hamiltonian which gives rise to the mean-field approximation (i.e., expression(\ref{H_single_wire})) is given by
\begin{eqnarray}
H_{full}=\sum_{i=1}^{N-1}-a^\dagger_{i}a_{i+1}+h.c.-4a^\dagger_ia^\dagger_{i+1}a_{i+1}a_i. \label{H_single_wire_full}
\end{eqnarray}\\

Next, let's try to move the left Majorana zero mode $\gamma_{1A}$ to the right by one site and in the meanwhile move the right Majorana zero mode $\gamma_{2B}$ to the right by one site in order to keep average particle number constant (see figure \ref{fig_single_wire}). Let the tuning mean-field Hamiltonian to be
\begin{eqnarray}
H'&=&2\lambda \mu_1(\lambda)a^\dagger_1a_1+(1-\lambda)(a^\dagger_1+a_1)(a^\dagger_2-a_2) \nonumber \\
&+&2(1-\lambda)\mu_{N+1}(\lambda)a^\dagger_{N+1}a_{N+1}+\lambda(a^\dagger_N+a_N)(a^\dagger_{N+1}-a_{N+1}). 
\label{tunning_single_wire}
\end{eqnarray}
(Note that the system Hamiltonian can be read off from a graphic representation such as that shown in figure \ref{fig_single_wire} (it represents $H+H'$ with $H$ and $H'$ given by equation (\ref{H_single_wire}) or equivalently equation (\ref{H_single_wire_MF}) and equation (\ref{tunning_single_wire}), respectively) according to the following rules. Each site is represented by two dots corresponding to real and imaginary part of the fermion on the site. The left dot represents $\gamma_{i,A}$ and the right represents $\gamma_{i,B}$ and the fermion creation operator one the site $i$ is $a^\dagger_i=1/2(\gamma_{i,A}+i\gamma_{i,B})$. Depending on direction, each link with arrow represents either $i\gamma_{i,B}\gamma_{i+1,A}$ (pointing from site $i$ to site $i+1$) or $i\gamma_{i+1,A}\gamma_{i,B}$ (pointing from $i+1$ to $i$). A link with varying strength parameterized by $\lambda$ is represented by a dashed line. An isolated dot represents an unoccupied site with positive chemical potential on it.) We require that as $\lambda$ goes from 0 to 1, both Majorana zero modes are moved to the right by one site. This is easily achieved as long as $\mu_1(\lambda)>0$ and $\mu_{N+1}(\lambda)>0$. As shown in figure \ref{fig_single_wire}, at stage (a), there is one Majorana zero mode $\gamma_{1,A}$ sitting at site 1, and one Majorana zero mode $\gamma_{N,B}$ sitting at site N. Site $N+1$ is initially unoccupied.  At intermediate stage (b), the strength of the link (i.e., the magnitude of coefficient of $\gamma_{1,B}\gamma_{2,A}$ of the Hamiltonian) between site $1$ and $2$ decreases whereas the strength of the link between $N$ and $N+1$ increases (remember that the changing strength of links is represented by dashed lines), meanwhile the chemical potential on site 1 increases and that on site $N+1$ decreases. At final stage (c), site 1 is unoccupied and the link between site 1 and 2 vanishes, whereas strength of link between site $N$ and $N+1$ reaches final value. The Majorana zero modes are now sitting at site 2 and $N+1$, respectively. Constant average particle number during the process can be achieved by tuning functions $\mu_1(\lambda)$ and $\mu_{N+1}(\lambda)$. Explicit calculation shows that the order parameters  $\langle a_{i+1}a_i\rangle$ for $i=2,\cdots,N-1$ remain unchanged while for $i=1$ and $i=N$, they change as functions of $\lambda$. The gap for $i=1$ and $i=N$ can be tuned to be equal to the value we need by tuning the corresponding nearest neighbor interactions $V_{i,i+1}$ since the gap for link $i-i+1$ is equal to $V_{i,i+1}\langle a_{i+1}a_i\rangle$. So both criteria are satisfied in the process of moving Majorana zero modes and at the same time we are able to calculate explicitly the many-body wave functions of the system. In Section \ref{Double} and \ref{Single}, both criteria will be checked explicitly throughout the braiding process. \\
\vspace{50pt}
\begin{figure}[h!]
\begin{center}
\includegraphics[width=0.6\textwidth]{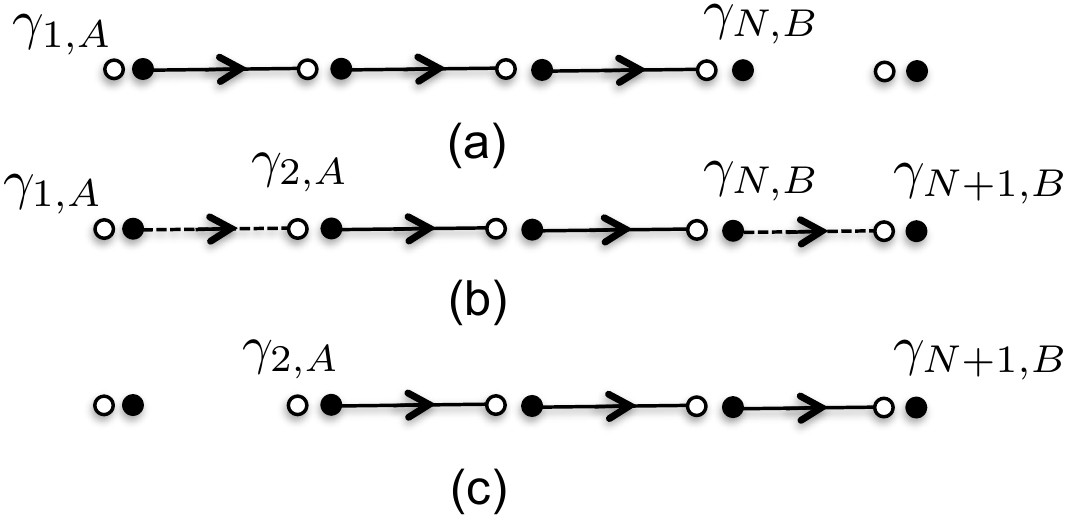}
\end{center}
\vspace{20pt}
\caption{Moving Majorana zero modes in a single wire. Black and white dots represent two Majorana fermions at each site. (a) initial configuration. (b) Moving both the left and right Majorana zero modes to the right by one site. (c) Final configuration. }\label{fig_single_wire}
\end{figure}

\subsection{Non-Abelian transformation and gauge invariance} \label{Non-Abelian}

Consider a Hamiltonian $H(\lambda)$ depending continuously on $\lambda$ with $n$ degenerate levels which do not cross other levels. As $H(\lambda)$ is adiabatically varied and returned to the initial one, a set of $n$ states which at time $t_i$ are degenerate orthonormal eigenstates of $H(\lambda_i)$ undergo non-abelian transformation and each of the final states is a unitary transform of the initial states. For an arbitrary smooth set of bases $\psi_a(t)$, the solutions to the time-dependent Schrodinger equations $\eta_a(t)$ can be written as
\begin{eqnarray}
\eta_a(t)=U_{ab}(t)\psi_b(t) \label{unitary_trans}
\end{eqnarray}
with the initial condition $\eta_a(t_i)=\psi_a(t_i)$, $a=1,2,\cdots, n$. $U_{ab}(t)$ is found  to be \cite{WZ}
\begin{eqnarray}
U(t)=P\mathrm{exp} \int_{t_i}^tA(\tau)d\tau, \label{U}
\end{eqnarray}
where Berry connection $A$ is given by
\begin{eqnarray}
A_{ab}=(\psi_a,\dot{\psi}_b)^* \label{A}
\end{eqnarray}
with $\dot{\psi}_b\equiv\partial\psi_b/\partial\tau$ and $P$ is the time-ordering operator. \\

If we choose another set of basis $\tilde{\psi}(t)=\Omega(t)\psi(t)$, $A$ transforms as 
\begin{eqnarray}
\tilde{A}=\dot{\Omega}\Omega^{-1}+\Omega A\Omega^{-1}, \label{tilde_A}
\end{eqnarray}
and $U$ transforms as
\begin{eqnarray}
\tilde{U}=\Omega(t_i)U\Omega^{-1}(t). \label{tilde_U}
\end{eqnarray}
The solutions to the time-dependent Schrodinger equations $\tilde{\eta}_a(t)$ become
\begin{eqnarray}
\tilde{\eta}_a(t)=\tilde{U}_{ab}(t)\tilde{\psi}_b(t)=\Omega_{ab}(t_i)\eta_b(t) \label{tilde_unitary}
\end{eqnarray}
with initial condition $\tilde{\eta}_a(t_i)=\tilde{\psi}_a(t_i)=\Omega_{ab}(t_i)\psi_b(t_i)$. Equation (\ref{tilde_unitary}) demonstrates gauge invariance of the evolution of the solutions, i.e., the evolution is independent of different choices of $\Omega$ so that the evolution of ground states is uniquely determined by their initial states (in equation (\ref{tilde_unitary}), the time evolution of $\tilde{\eta}_a(t)$ is independent of time dependence of $\Omega_{ab}(t)$ and is completely determined by the initial values $\Omega_{ab}(t_i)$). In the following two sections, we shall calculate evolution of degenerate ground states in different choices of bases to confirm the gauge independence and so to justify our choice of basis for calculating Berry phase in \ref{Single}. \\

\subsection{Double interchange of Majorana zero modes} \label{Double}

In this and the next sections, we'll consider systems harboring four Majorana zero modes $\gamma_i$, $i=1,2,3,4$ and interchange $\gamma_1$ and $\gamma_2$. It's simplest to discuss braiding in the diagonal basis: such a basis is given by
$|00\rangle$ and $|11\rangle=f^\dagger_1f^\dagger_2|00\rangle$, $f_1=\gamma_1+i\gamma_2$ and $f_2=\gamma_3+i\gamma_4$ and we interchange positions of $\gamma_1$ and $\gamma_2$. In this basis, the off-diagonal Berry connections vanish by construction. The off-diagonal Berry connection $\langle00|\dot{11}\rangle$ ( ' $\dot{}$ ' denotes derivative) is
\begin{eqnarray}
\langle00|\dot{11}\rangle=\langle00|\dot{f}^\dagger_1f^\dagger_2|00\rangle+\langle00|f^\dagger_1f^\dagger_2|\dot{00}\rangle \label{Berry_0011_diag}
\end{eqnarray}
Both terms that contribute to $\langle00|\dot{11}\rangle$ vanish: for the first term, we can first switch positions of $\dot{f}^\dagger_1$ and $f^\dagger_2$ with a minus sign since they anti-commute due to Fermi statistics and their exponentially small spatial overlap and next operate $f^\dagger_2$ on $\langle00|$ giving zero by definition since $f_2|00\rangle=0$; for the second term, $\langle00|f^\dagger_1=0$ by definition. Thus $\langle00|\dot{11}\rangle=0$. In this basis, if we can make $|00\rangle$ and $|11\rangle$ real throughout the braiding, the Berry phase vanishes for each state. Then the braiding is completely determined by the explicit change (monodromy) of the basis states. This can be realized in the set up shown in figure \ref{fig_braiding}. The arrows and links have the same meaning as discussed below equation (\ref{tunning_single_wire}) in Section \ref{Criteria Check}. When a link goes from solid to dotted, it means the strength of the link decreases continuously to zero (and vice versa). An isolated dot denotes an unoccupied site. When the dot goes from isolated to connected by a link, the chemical potential on that site decreases to zero (and vice versa). According the rules given below equation (\ref{tunning_single_wire}), one can easily read off the corresponding Hamiltonian for the braiding process from figure \ref{fig_braiding}.  \\

\begin{figure}[h!]
\begin{center}
\includegraphics[width=0.8\textwidth]{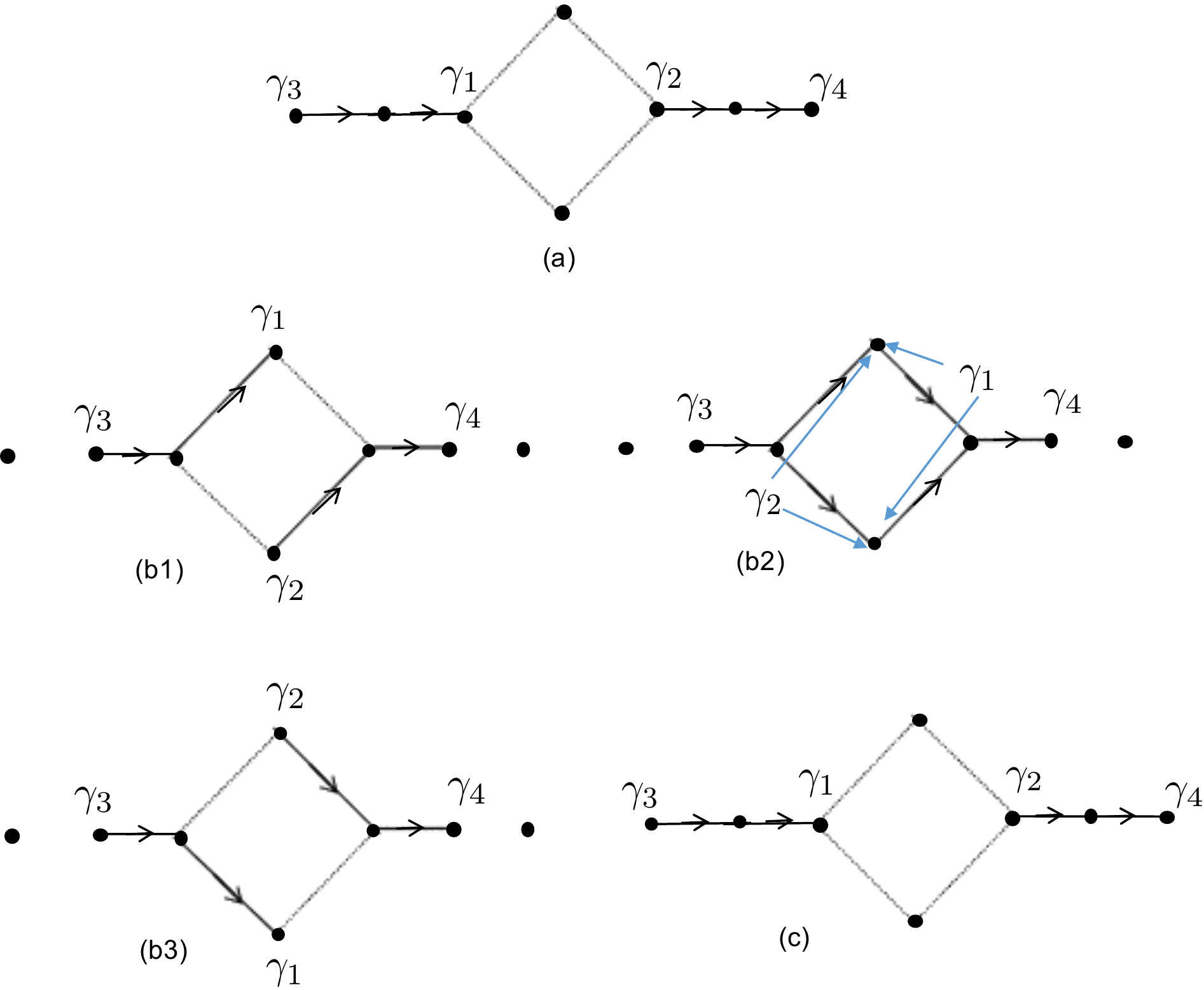}
\end{center}
\vspace{20pt}
\caption{Exchanging Majorana zero modes $\gamma_1$ and $\gamma_2$ twice in 1D Kitaev network. (a) Initial configuration. (b1) - (b3) Intermediate  configuration. At (b2), $\gamma_1$ is localized at both top and bottom sites in the figure and so is $\gamma_2$; the former are localized on the right of the top and bottom sites (i.e., imaginary part of fermion, $\gamma_B$ in the sense of figure \ref{fig_single_wire}) and the latter are localized on the left (i.e., real part, $\gamma_A$). For simplicity, we didn't draw each site by two dots as we did in figure \ref{fig_single_wire}. If we were to do that, then $\gamma_1$ is localized at the right dot on the top and bottom site and $\gamma_2$ is at the left dot on the top and the bottom site. The other two Majorana zero modes $\gamma_3$ and $\gamma_4$ are also moved, but they eventually go back to their initial positions and very importantly they never cross other Majorana zero modes. (c) Final configuration.}\label{fig_braiding}
\end{figure}

Adiabatic evolution of the Hamiltonian according to figure \ref{fig_braiding} yields double interchange of $\gamma_1$ and $\gamma_2$ at the end of the braiding and both of the operators continuously evolve back to themselves with extra minus sign (note that this result is the same as that with interchanging two vortices and therefore two Majorana zero modes twice in p+ip superfluids). Therefore $|11\rangle$ picks up an explicit phase of $\pi$ relative to $|00\rangle$ at the end of the braiding. Combining with the (trivial) Berry phase contribution, the final states evolve to $|\eta(00)\rangle_f=e^{i\phi}|\eta(00)\rangle_i$, $|\eta(11)\rangle_f=e^{i(\phi+\pi)}|\eta(11)\rangle_i$ with initial condition $|\eta(00)\rangle_i=|00\rangle_i$ and $|\eta(11)\rangle_i=|11\rangle_i$ (we use $\eta$ to denote actual solutions of time-dependent Schrodinger equations). Now if we switch to another basis $|\tilde{00}\rangle=\alpha|00\rangle+\beta|11\rangle$ and $|\tilde{11}\rangle=\beta^*|00\rangle-\alpha^*|11\rangle$, with initial condition $|\eta(\tilde{00})\rangle_i=|\tilde{00}\rangle_i$ and $|\eta(\tilde{11})\rangle_i=|\tilde{11}\rangle_i$ (assume constant $\alpha$ and $\beta$), they should evolve to
\begin{eqnarray}
|\eta(\tilde{00})\rangle_f=\alpha|\eta(00)\rangle_f+\beta|\eta(11)\rangle_f=e^{i\phi}(\alpha |\eta(00)\rangle_i-\beta|\eta(11)\rangle_i) \nonumber \\
|\eta(\tilde{11})\rangle_f=\beta^*|\eta(00)\rangle_f-\alpha^*|\eta(11)\rangle_f=e^{i\phi}(\beta^*|\eta(00)\rangle_i+\alpha^*|\eta(11)\rangle_i).\label{tilde_trans}
\end{eqnarray}
Each of them evolves into linear combination of the initial states. The $U$ matrix in this basis should be identity according to equation (\ref{tilde_U}) since the in the diagonal basis, $U$ is identity. This can be easily verified. The Berry connection matrix elements are
\begin{eqnarray}
\langle\tilde{00}|\dot{\tilde{00}}\rangle&=&2\mathrm{Im}(\alpha^*\beta)\langle 00|\dot{11}\rangle \nonumber \\
\langle\tilde{11}|\dot{\tilde{11}}\rangle&=&-\langle\tilde{00}|\dot{\tilde{00}}\rangle \nonumber \\
\langle\tilde{00}|\dot{\tilde{11}}\rangle&=&-\mathrm{Im}(\alpha^{*2}+\beta^{*2})\langle00|\dot{11}\rangle, \label{Berry_tilde_double}
\end{eqnarray}
which according to equation (\ref{Berry_0011_diag}) all vanish. \\

So far, we have discussed particle-number non-conserving states. What about particle-number conserving states? Equation (\ref{Berry_0011_diag}) is not strictly satisfied for finite size systems. In the thermodynamic limit, it is satisfied. The Berry phase associated with each basis state still vanishes since they are real after particle number projection. So in this braiding scheme, the braiding statistics is unchanged from particle-number non-conserving case. \\

To verify the above result, we have performed explicit calculations in the diagonal basis with average number of 4 electrons for particle non-conserving states (throughout the braiding, the hopping and gap parameter are set equal). Our calculations confirm that the two ground states evolve continuously as we expect. The two criteria listed in Section \ref{Criteria Check} are satisfied and the energy gap never closes during the braiding. There is one caveat: at the stage where the two Majorana fermions interchange places (cf. (b2) in figure \ref{fig_braiding}), the average particle number of $|11\rangle$ differs from that of $|00\rangle$ with a maximum discrepancy of one electron in the middle of the stage. Although this issue is absent for some basis states, constant average particle number can not be satisfied for states in all bases. In practice, we may ensure the constant average particle number in one basis state and obtain the particle conserving state for the other basis state by applying particle number conserved BdG operators to the former state. \\

\subsection{Single interchange of Majorana zero modes} \label{Single}

It's more interesting to know the braiding statistics arising from interchanging two Majorana zero modes once. This can be realized in the T-junction geometry proposed by Alicea et al \cite{Alicea}. It turns out that braiding in particle-number conserving systems can also be realized in the same T-junction provided we can ensure constant average particle number by, for example, moving the other two zero Majorana fermions accordingly. This can be achieved similarly to the double interchange scheme. The braiding process is shown in figure \ref{fig_single_braiding} where we have adopted a simplified configuration which can realize Majorana zero modes interchange as a T-junction does. The arrows, links and isolated dots have the same meaning as discussed in the above section and the corresponding Hamiltonian can be easily read off from figure \ref{fig_single_braiding} according to the rules laid out below equation (\ref{tunning_single_wire}). \\

\begin{figure}[h!]
\begin{center}
\includegraphics[width=0.8\textwidth]{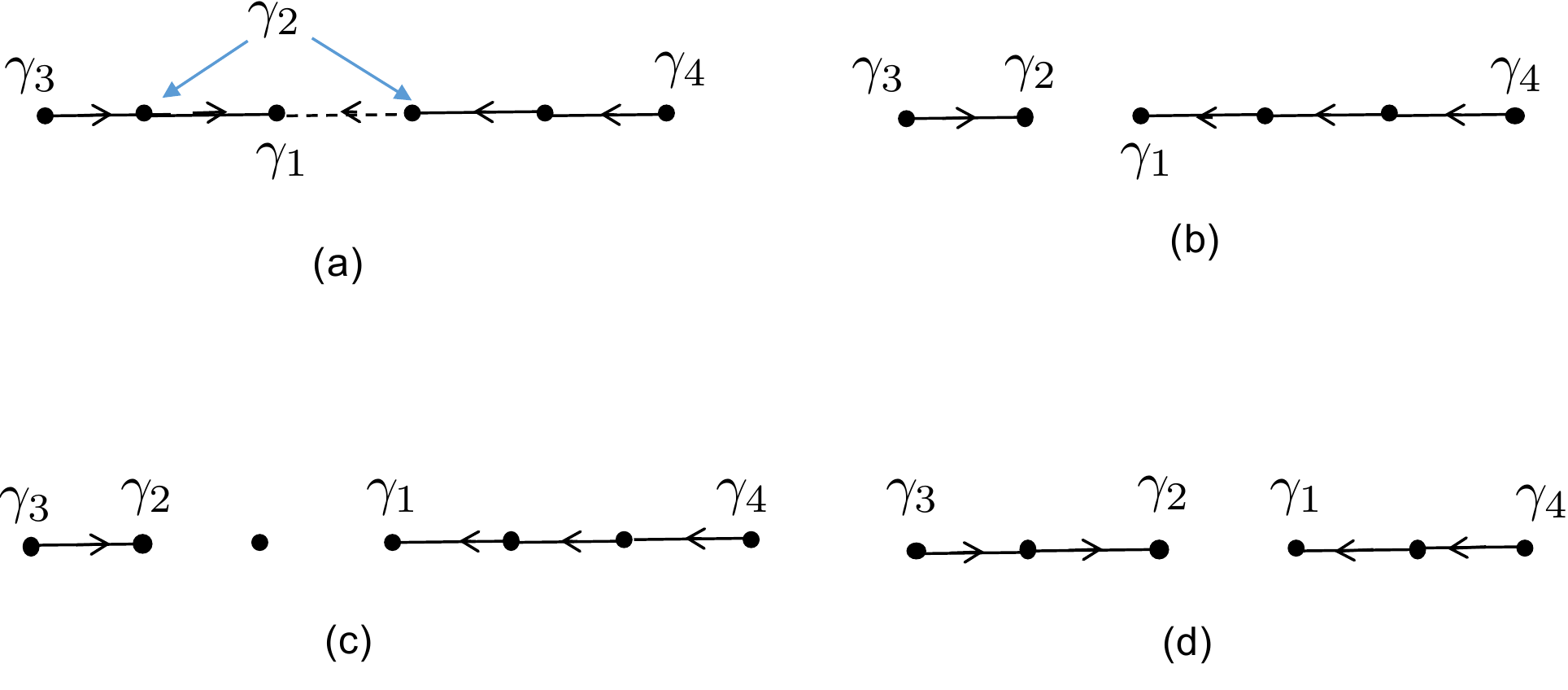}
\end{center}
\vspace{30pt}
\caption{Exchanging Majorana zero modes $\gamma_1$ and $\gamma_2$ in a 1D Kitaev network. (a) $\gamma_2$ is moved to the left as indicated by the two blue arrows; in the process of moving, $\gamma_2$ is localized at two sites as indicated by the two blue arrows. (b) $\gamma_1$ and $\gamma_2$ are interchanged. (c) $\gamma_1$ is moved to the right to the original place of $\gamma_2$ before interchange. (d) $\gamma_2$ is moved to the right to the original place of $\gamma_1$ before interchange. The final configuration is the same as before interchange with $\gamma_1$ and $\gamma_2$ having swapped places. From (b) to (c) and from (c) to (d), $\gamma_4$ is also moved to ensure particle number conservation. }\label{fig_single_braiding}
\end{figure}

In the diagonal basis, the two basis functions are no longer real. In the particle-number non-conserving form, the two states have the same Berry phase during the interchange. However, it is not necessarily true for the particle-number conserving states and we may expect different braiding statistics for particle-number conserving states! Switching to another basis with real basis states doesn't alter the conclusion due to gauge invariance. Let us check gauge invariance by starting from a non-diagonal basis given by $|00\rangle$ and $|11\rangle=f^\dagger_1f^\dagger_2|00\rangle$ with $f_1=\gamma_3+i\gamma_1$ and $f_2=\gamma_4+i\gamma_2$ which are real throughout the interchange process. At the end of the braiding, $\gamma_1$ becomes $\pm\gamma_2$ and $\gamma_2$ becomes $\mp\gamma_1$. We haven't specified the signs after the braiding which depend on the braiding sequence (clockwise or counterclockwise). At the end of the braiding, the basis functions become (according to the definition that $|11\rangle=f^\dagger_1f^\dagger_2|00\rangle$ with $f_1=\gamma_3+i\gamma_1$ and $f_2=\gamma_4+i\gamma_2$)

\begin{eqnarray}
|00\rangle_f&=&\frac{1}{\sqrt{2}}(|00\rangle_i\pm|11\rangle_i) \nonumber \\
|11\rangle_f&=&\frac{1}{\sqrt{2}}(|11\rangle_i\mp|00\rangle_i). \label{mono_off_diag}
\end{eqnarray}
The transformation matrix $U$ for the solutions to the time-dependent Schrodinger equations is found to be (cf. equation (\ref{U}))
\begin{eqnarray}
U=\left(\begin{array}{cc} \mathrm{cos}\theta& \mathrm{sin}\theta \\ -\mathrm{sin}\theta  & \mathrm{cos}\theta \end{array}\right) \label{U_off}
\end{eqnarray}
with $\theta=\int_{t_i}^{t_f}dt \langle00(t)|\dot{11}(t)\rangle$.
Combining equation (\ref{U_off}) with (\ref{mono_off_diag}), we get the solutions at the end of braiding 
\begin{eqnarray}
\left(\begin{array}{c} |\eta(00)\rangle_f \\ |\eta(11)\rangle_f \end{array}\right) &=&\frac{1}{\sqrt{2}}\left(\begin{array}{cc} \mathrm{cos}\theta& \mathrm{sin}\theta \\ -\mathrm{sin}\theta  & \mathrm{cos}\theta \end{array}\right) \left(\begin{array}{cc} 1& \pm1 \\ \mp 1  & 1 \end{array}\right)\left(\begin{array}{c} |\eta(00)\rangle_i \\ |\eta(11)\rangle_i \end{array}\right) \nonumber \\
&=&\left(\begin{array}{cc} \mathrm{cos}(\theta\pm\frac{\pi}{4})& \mathrm{sin}(\theta \pm \frac{\pi}{4})\\ -\mathrm{sin}(\theta\pm \frac{\pi}{4}  & \mathrm{cos}(\theta\pm \frac{\pi}{4}) \end{array}\right)\left(\begin{array}{c} |\eta(00)\rangle_i \\ |\eta(11)\rangle_i \end{array}\right). \label{braiding_off_diag}
\end{eqnarray}

Now, we switch back to the diagonal basis given by $|\tilde{00}\rangle$ and $|\tilde{11}\rangle=\tilde{f}^\dagger_1\tilde{f}^\dagger_2|\tilde{00}\rangle$ with $\tilde{f}_1=\gamma_3+i\gamma_4$ and $\tilde{f}_2=\gamma_1-i\gamma_2$. The diagonal basis states are related to the non-diagonal basis states by
\begin{eqnarray}
|\tilde{00}\rangle&=&\frac{1}{\sqrt{2}}(|00\rangle-i|11\rangle) \nonumber \\
|\tilde{11}\rangle&=&\frac{1}{\sqrt{2}}(|00\rangle+i|11\rangle). \label{basis_trans}
\end{eqnarray}
At the end of the braiding, they are
\begin{eqnarray}
|\tilde{00}\rangle_f&=&e^{\pm i\pi/4}(|\tilde{00}\rangle_i \nonumber \\
|\tilde{11}\rangle_f&=&e^{\mp i\pi/4}|\tilde{11}\rangle_i. \label{mono_diag}
\end{eqnarray}
The U matrix is
\begin{eqnarray}
U=\left(\begin{array}{cc} e^{i\theta}& 0\\ 0 & e^{-i\theta} \end{array}\right). \label{U_diag}
\end{eqnarray}
From equation (\ref{basis_trans}) and (\ref{braiding_off_diag}), we obtain the final solutions in the diagonal basis
\begin{eqnarray}
\left(\begin{array}{c} |\eta(\tilde{00})\rangle_f \\ |\eta(\tilde{11})\rangle_f \end{array}\right) &=&\frac{1}{\sqrt{2}}\left(\begin{array}{cc} 1& -i\\ 1  & i \end{array}\right)\left(\begin{array}{c} |\eta(00)\rangle_f \\ |\eta(11)\rangle_f \end{array}\right)  \nonumber \\
&=&\frac{1}{\sqrt{2}}\left(\begin{array}{cc} 1& -i\\ 1  & i \end{array}\right)\left(\begin{array}{cc} \mathrm{cos}(\theta\pm\frac{\pi}{4})& \mathrm{sin}(\theta \pm \frac{\pi}{4})\\ -\mathrm{sin}(\theta\pm \frac{\pi}{4}  & \mathrm{cos}(\theta\pm \frac{\pi}{4}) \end{array}\right)\left(\begin{array}{c} |\eta(00)\rangle_i \\ |\eta(11)\rangle_i \end{array}\right) \nonumber \\
&=&\frac{1}{\sqrt{2}}\left(\begin{array}{cc} 1& -i\\ 1  & i \end{array}\right)\left(\begin{array}{cc} \mathrm{cos}(\theta\pm\frac{\pi}{4})& \mathrm{sin}(\theta \pm \frac{\pi}{4})\\ -\mathrm{sin}(\theta\pm \frac{\pi}{4}  & \mathrm{cos}(\theta\pm \frac{\pi}{4}) \end{array}\right)\frac{1}{\sqrt{2}}\left(\begin{array}{cc} 1& 1\\ i  & -i \end{array}\right)\left(\begin{array}{c} |\eta(\tilde{00})\rangle_i \\ |\eta(\tilde{11})\rangle_i \end{array}\right)  \nonumber \\
&=&\left(\begin{array}{cc} e^{i(\theta\pm\frac{\pi}{4})}& 0\\ 0 & e^{-i(\theta\pm\frac{\pi}{4})} \end{array}\right)\left(\begin{array}{c} |\eta(\tilde{00})\rangle_i \\ |\eta(\tilde{11})\rangle_i \end{array}\right). \label{braid_diag}
\end{eqnarray}
This is the same as obtained by combining equation (\ref{mono_diag}) and (\ref{U_diag}) 
\begin{eqnarray}
\left(\begin{array}{c} |\eta(\tilde{00})\rangle_f \\ |\eta(\tilde{11})\rangle_f \end{array}\right) &=&\left(\begin{array}{cc} e^{i\theta}& 0\\ 0 & e^{-i\theta} \end{array}\right) \left(\begin{array}{cc} e^{\pm i\pi/4}& 0\\ 0 & e^{\mp i\pi/4} \end{array}\right) \left(\begin{array}{c} |\tilde{00}\rangle_i \\ |\tilde{11}\rangle_i \end{array}\right) \nonumber \\
&=&\left(\begin{array}{cc} e^{i(\theta\pm\frac{\pi}{4})}& 0\\ 0 & e^{-i(\theta\pm\frac{\pi}{4})} \end{array}\right)\left(\begin{array}{c} |\eta(\tilde{00})\rangle_i \\ |\eta(\tilde{11})\rangle_i \end{array}\right). \label{braid_diag_gauge}
\end{eqnarray}

So consistency is satisfied. If $\theta$ is nonzero, then we get a different braiding result for particle-number conserving states compared to particle-number non-conserving states for which $\theta=0$.\\

We have implemented calculations for eight lattice sites and on average four particles. Calculations are done in an off-diagonal basis in which basis states are kept real during the interchange. The off-diagonal Berry phase $\theta$ (cf. the matrix in equation (\ref{U_off})) is calculated. We found that for particle-number conserving states with four particles, $\theta=0$ consistent with result for particle number non-conserving states. Interestingly, $\theta=\pm0.4$ for states in two and six particle number sectors respectively. It will be interesting to see how nonzero Berry phases away from average particle number scale with the system size. \\

\section{Derivation of Majorana wave functions in the alternative interchange process} \label{proof}

In this appendix, we derive the Majorana wave functions given by equation (\ref{u_case2}). We first obtain the solutions to the BdG equations in the rotated frame and then express the solutions in the original frame (see figure \ref{fig_vortex2_rotate}). In the rotated frame, the Majorana wave functions are independent of $\theta_0$ since they don't change their configuration in the rotated frame. So their wave functions in the rotated frame are given by their initial ones
\begin{eqnarray} 
u'_1(r,\theta_0)&=&u_1(r,\theta_0=0)=\mathrm{exp}\{\frac{\pi}{2} i\}u(|\vec{r}-\vec{R}_1|)e^{i\theta(\vec{r}-\vec{R}_1)} \nonumber \\
u'_2(r,\theta_0)&=&u_2(r,\theta_0=0)=u(|\vec{r}-\vec{R}_2|)e^{i\theta(\vec{r}-\vec{R}_2)}, \label{u'_case2}
\end{eqnarray}
where prime superscripts are used to denote solutions in the rotated frame; the factor $\mathrm{exp}\{\frac{\pi}{2} i\}$ of $u_1(r,\theta_0=0)$ comes from the overall phase of $\pi$ at the core of vortex 1 due to vortex 2 (cf. figure \ref{fig_vortex2_rotate}). Since the rotated frame is rotated by $\theta_0$ relative to the lab frame, the solutions in the lab frame will pick up a phase factor $\mathrm{exp}\{-\theta_0i\}$, yielding the expression in (\ref{u_case2}). Alternatively, we can compare the BdG equations in the two reference frames. The diagonal terms of the BdG equations take the same form, whereas the off-diagonal terms differ by a phase factor $\mathrm{exp}\{-2\theta_0i\}$ because $\partial_{x'}+i\partial_{y'}=(\partial_x+i\partial_y)\mathrm{exp}\{-\theta_0i\}$ and the center-of-mass degree of freedom of the gap in the rotated frame $\bigtriangleup'$ is related to that in the lab frame $\bigtriangleup$ by $\bigtriangleup'=\bigtriangleup\mathrm{exp}\{-\theta_0i\}$. So when transformed to the lab frame, $u_1$ and $u_2$ pick up an extra phase factor $\mathrm{exp}\{-\theta_0i\}$. End of derivation. \\

\section{The equivalence of the two interchange processes} \label{equiv}

In this appendix, we verify the equivalence of the two interchange processes discussed in Section \ref{braid}. Let's compare the ansatz for the Cooper pair wave function (\ref{Cooper_case2}) with (\ref{Cooper_case1}). The ansatz (\ref{Cooper_case2}) yields an order parameter center-of-mass phase which is smaller than the ansatz (\ref{Cooper_case1}) by $2\theta_0$. This agrees with the center-of-mass phase difference of the superfluid order parameter in the two interchanging processes. Notice that the total phase difference of the Cooper pair wave function between the two processes is $3\theta_0$, with a $2\theta_0$ contribution from the center of mass gap phase and a $\theta_0$ contribution from the relative gap phase difference, both of which appear in the BdG equations (so the overall phases of the gap  in the BdG equations in the two processes differ by $3\theta_0$). In addition, we have seen that the two interchanging processes yield the same braiding phase. It's thus tempting to ascribe the difference in the two processes to an overall phase factor. This overall phase factor may be regarded as a gauge choice, i.e., choice of instantaneous ground states, and we may conclude that the two processes belong to the same physical process. Let's examine the argument in more detail. Let's assume that $|0(\theta_0)\rangle_{2N}$ in the alternative process is the same as that in the standard one, with an extra phase factor of $\mathrm{exp}(-i3\theta_0N/2)$ (compare equation (\ref{Cooper_case2}) with (\ref{Cooper_case1})). If at the same time, the corresponding BdG operators in the two processes were the same up to a phase factor, then we can conclude that the two processes are identical. With the definition of the Majorana fermion operator (\ref{gamma_pc}), the Majorana wave functions in the two processes (\ref{u_case1}), (\ref{u_case2}) and the assumption that  $C^\dagger_{II}(\theta_0)=C^\dagger_I(\theta_0)\mathrm{exp}(-i3\theta_0 )$ (Roman subscripts refer to the two processes), we can explicitly write down the expressions for the BdG operators in the two processes
\begin{eqnarray}
\alpha_I^\dagger(\theta_0)&=& \int d^2r \hspace{5pt} \mathrm{exp}\{\frac{\theta_0 i}{2}\}(u(|\vec{r}-\vec{R}_1|)e^{i(\theta(\vec{r}-\vec{R}_1)+\pi/2)} \nonumber+iu(|\vec{r}-\vec{R}_2|)e^{i\theta(\vec{r}-\vec{R}_2)})\psi^\dagger(r)  \nonumber \\
&+&\mathrm{exp}\{-\frac{\theta_0 i}{2}\}(u(|\vec{r}-\vec{R}_1|)e^{-i(\theta(\vec{r}-\vec{R}_1)+\pi/2)} \nonumber+iu(|\vec{r}-\vec{R}_2|)e^{-i\theta(\vec{r}-\vec{R}_2)})\psi(r)C_I^\dagger(\theta_0), \nonumber \\ 
& &\ \nonumber \\
\alpha_{II}^\dagger(\theta_0)&=& \int d^2r \hspace{5pt} \mathrm{exp}\{-\theta_0 i\}(u(|\vec{r}-\vec{R}_1|)e^{i(\theta(\vec{r}-\vec{R}_1)+\pi/2)} \nonumber+iu(|\vec{r}-\vec{R}_2|)e^{i\theta(\vec{r}-\vec{R}_2)})\psi^\dagger(r)  \nonumber \\
&+&\mathrm{exp}\{-2\theta_0i\}(u(|\vec{r}-\vec{R}_1|)e^{-i(\theta(\vec{r}-\vec{R}_1)+\pi/2)} \nonumber+iu(|\vec{r}-\vec{R}_2|)e^{-i\theta(\vec{r}-\vec{R}_2)})\psi(r)C_I^\dagger(\theta_0). \label{alpha_I_II} 
\end{eqnarray}
We see that $\alpha^\dagger_{II}(\theta_0)$ is proportional to $\alpha^\dagger_I(\theta_0)$, with a phase factor $\mathrm{exp}\{-3\theta_0/2i\}$. So we may indeed identify the two processes!\\

\section{Particle number conservation in the BdG formalism}  \label{continuity_BdG}

In this appendix, we discuss particle number conservation in the BdG formalism. As a BdG Hamiltonian doesn't conserve particle number, it is worth asking whether the many-body eigenstates corresponding to the BdG equations satisfy particle number conservation for the BdG Hamiltonian. This question has been discussed by several authors {\cite{Furusaki, Bagwell,Sols}}. I shall follow \cite{Sols} in which contributions to charge and current from condensate and quasiparticles are clearly distinguished, which is useful for our discussions. We will discuss particle number conservation (or lack of it) associated with the BdG equations in general inhomogeneous s-wave superconducting states whose quasiparticle energies are degenerate  in spin (i.e. no spin-dependent potential). Following this, we will show how a straightforward modification within the BdG formalism can recover particle number conservation for a general excited many-body eigenstate.\\

Consider the BdG equations  

\begin{eqnarray}
\left(\begin{array}{cc}H_0& \bigtriangleup \\ \bigtriangleup^* & -H_0^*\end{array}\right)\left(\begin{array}{c}u_n \\v_n\end{array}\right) 
&=& \epsilon_n \left(\begin{array}{c}u_n \\v_n \end{array}\right) \label{BdG}
\end{eqnarray}
where $H_0$ is the single particle Hamiltonian (spin-independent), $\bigtriangleup$ is the gap function and $(u_n(r),v_n(r))$ and $\epsilon_n>0$ are the normalized wave functions and the eigen-energies, respectively. The corresponding BdG Hamiltonian can be written as
\begin{eqnarray}
H=-\sum_{n\sigma}\epsilon_n\int dr |v_n(r)|^2+\sum_{n\sigma}\epsilon_n\gamma^\dagger_{n\sigma}\gamma_{n\sigma} \label{BdG_H}
\end{eqnarray}
where $\gamma^\dagger_{n\sigma}$ creates the quasiparticle $n$ with spin $\sigma$. 
We now express the charge and current density operators in terms of the quasiparticle operators 
\begin{eqnarray}
\rho&=&e(\sum_{n\sigma}|v_n|^2+\sum_{nm\sigma}(u^*_nu_m-v^*_nv_m)\gamma^\dagger_{n\sigma}\gamma_{m\sigma}+\sum_{nm\sigma}u_nv_m\sigma\gamma_{m\sigma}\gamma_{n,-\sigma}+\sum_{nm\sigma}u^*_nv^*_m\sigma\gamma^\dagger_{n,-\sigma}\gamma^\dagger_{m\sigma}) \nonumber \\
j&=&\frac{e}{2mi}(-\sum_{n\sigma}v^*_nDv_n+\sum_{nm\sigma}(u^*_nDu_m+v^*_nDv_m)\gamma^\dagger_{n\sigma}\gamma_{m\sigma}+\sum_{nm\sigma}v_mDu_n\sigma\gamma_{m\sigma}\gamma_{n,-\sigma}-\sum_{nm\sigma}v^*_mDu^*_n\sigma\gamma^\dagger_{n,-\sigma}\gamma^\dagger_{m\sigma}) \nonumber \\
\label{dens-current}
\end{eqnarray}
where $e=-|e|$, $D$ is defined as $fDg\equiv f\bigtriangledown g -(\bigtriangledown f)g$. In the above equation, the contribution from the condensate (the first term in $\rho$ and $j$) is separated from that from the quasiparticles. \\

Let's now consider particle number conservation within the BdG formalism. The time-derivative of the density operator is obtained from equations (\ref{BdG_H}) and (\ref{dens-current})
\begin{eqnarray}
\dot{\rho}&=&\frac{1}{i}[\rho,H] \nonumber \\
&=&\frac{e}{i}\sum_{nm\sigma}((\epsilon_n-\epsilon_m)(v^*_nv_m-u^*_nu_m)\gamma^\dagger_{n\sigma}\gamma_{m\sigma}+(\epsilon_n+\epsilon_m)\sigma(u_nv_m\gamma_{m\sigma}\gamma_{n,-\sigma}+u^*_nv^*_m\gamma^\dagger_{n,\sigma}\gamma^\dagger_{m,-\sigma})) \label{drho_dt}
\end{eqnarray}
which yields $\langle\dot{\rho}\rangle=0$ for an eigenstate of the BdG Hamiltonian which is also an eigenstate of BdG quasiparticle occupation number and for thermal equilibrium satisfying $\langle\gamma^\dagger_{n\sigma}\gamma_{n'\sigma'}\rangle=f_n\delta_{nn'}\delta_{\sigma\sigma'}$ and $\langle\gamma_{n\sigma}\gamma_{n'\sigma'}\rangle=0$, where $f_n$ is the average occupation number of the quasiparticle. The divergence of current is obtained from equations (\ref{dens-current}) and (\ref{BdG})
\begin{eqnarray}
\bigtriangledown\cdot j&=&2e\mathrm{Im}(\bigtriangleup^*\sum_{n\sigma}u_nv^*_n)+\frac{e}{i}\{ 2(\bigtriangleup\sum_{nm\sigma}v_nu^*_m-\bigtriangleup^*\sum_{nm\sigma}u_nv^*_m)\gamma^\dagger_{m\sigma}\gamma_{n\sigma} \nonumber \\
&+&\sum_{mn\sigma}(\epsilon_m-\epsilon_n)(u_nu^*_m-v_nv^*_m)\gamma^\dagger_{m,\sigma}\gamma_{n,\sigma}+
(\bigtriangleup\sum_{nm\sigma}v_nv_m+\bigtriangleup^*\sum_{nm\sigma}u_nu_m)\sigma\gamma_{m,\sigma}\gamma_{n,-\sigma} \nonumber \\
&-&\sum_{nm\sigma}u_nv_m(\epsilon_n+\epsilon_m)\sigma\gamma_{m,\sigma}\gamma_{n,-\sigma}-(\bigtriangleup^*\sum_{nm\sigma}v^*_nv^*_m+\bigtriangleup\sum_{nm\sigma}u^*_mu^*_n)\sigma\gamma^\dagger_{n,-\sigma}\gamma^\dagger_{m,\sigma} \nonumber \\
&+&\sum_{nm\sigma}u^*_nv^*_m(\epsilon_n+\epsilon_m)\sigma\gamma^\dagger_{n,-\sigma}\gamma^\dagger_{m,\sigma}\} \label{divJ}
\end{eqnarray}

For the ground state and for the thermal average at equilibrium, particle number conservation is satisfied

\begin{eqnarray}
\langle\bigtriangledown\cdot j\rangle&=&2e\sum_{n\sigma}\mathrm{Im}\{\bigtriangleup^*u_nv^*_n(1-2f_n)\} \nonumber \\
&=&\frac{2e}{V}\mathrm{Im}\{|\bigtriangleup|^2\}=0\label{current_thermal}
\end{eqnarray}
where the second line is derived using the self-consistent gap equation
\begin{eqnarray}
\bigtriangleup=V\sum_{n\sigma}u_nv^*_n(1-2f_n) \label{gap}
\end{eqnarray}

However, for an arbitrary state, particle number conservation is, in general, not necessarily satisfied in the BdG formalism. The particle number conservation condition is obtained by adding equations (\ref{drho_dt}) and (\ref{divJ})
\begin{eqnarray}
\dot{\rho}+\bigtriangledown\cdot{j}&=&2e\mathrm{Im}(\bigtriangleup^*\sum_{n\sigma}u_nv^*_n)-\frac{2e}{i}(\bigtriangleup^*\sum_{nm\sigma}u_nv^*_m\gamma^\dagger_{m\sigma}\gamma_{n\sigma}-\mathrm{h.c.}) \nonumber \\
&+&\frac{e}{i}((\bigtriangleup\sum_{nm\sigma}v_nv_m+\bigtriangleup^*\sum_{nm\sigma}u_nu_m)\sigma\gamma_{m\sigma}\gamma_{n,-\sigma}-\mathrm{h.c.}) \label{continuity}
\end{eqnarray}

In general, the rhs of (\ref{continuity}) is non-zero, violating particle number conservation. Let's, for example, consider an eigenstate with say, one excited BdG quasiparticle $(n,\uparrow)$, then equation (\ref{continuity}) is evaluated to be
\begin{eqnarray}
\langle\dot{\rho}+\bigtriangledown\cdot j\rangle_n=2e\mathrm{Im}\{\bigtriangleup^*(\sum_{m,\sigma\neq (n, \uparrow)}u_mv^*_m-u_nv^*_n)\} \label{continuity_n}
\end{eqnarray}
The rhs of equation (\ref{continuity_n}) doesn't vanish, since the gap equation for the ground state reads
\begin{eqnarray}
\bigtriangleup = V\sum_{n, \sigma}u_nv^*_n \label{gap_gs}
\end{eqnarray}
and equation (\ref{continuity_n}) becomes
\begin{eqnarray}
\langle\dot{\rho}+\bigtriangledown\cdot j\rangle_n=-4e\mathrm{Im}\{\bigtriangleup^*u_nv^*_n\} \label{continuity_nf} 
\end{eqnarray}
In most cases, violation of particle number conservation is of order $1/N$ ($N$: total number of particles). However, for a localized quasiparticle, the rhs of equation (\ref{continuity_nf}) can become non-negligible in the localized region. \\

We can recover particle number conservation for a general excited eigenstate within the BdG framework by simply changing the gap equation so that it corresponds to the excited eigenstate in consideration. If we change the gap equation (\ref{gap_gs}) to
\begin{eqnarray}
\bigtriangleup = V(\sum_{m,\sigma\neq (n, \uparrow)}u_mv^*_m-u_nv^*_n) \label{gap_ex}
\end{eqnarray}
equation (\ref{continuity_n}) becomes
\begin{eqnarray}
\langle\dot{\rho}+\bigtriangledown\cdot j\rangle_n= \frac{2e}{V}\mathrm{Im}\{|\bigtriangleup|^2\}=0 \label{continuity_mod} 
\end{eqnarray}
and the continuity condition is satisfied. Equation (\ref{gap_ex}) is just the gap equation for the excited eigenstate in which the quasiparticle $(n,\uparrow)$ is occupied. \\

\end{appendices}


\begin{thebibliography}{99}
\bibitem{Leggett} A. J. Leggett, S. Chakravarty, A. T. Dorsey, Matthew P. A. Fisher, Anupam Garg, and W. Zwerger, Rev. Mod. Phys. 59, 1 (1987)
\bibitem{Clarke} Clarke and Wilhelm, Nature, 453, 1031 (2008)
\bibitem{Kitaev_top_comp} A.Yu. Kitaev, Ann. Phys., Volume 303, Issue 1, January 2003, Pages 2-30; for a review of topological quantum computing, see Chetan Nayak, Steven H. Simon, Ady Stern, Michael Freedman and Sankar Das Sarma, Rev. Mod. Phys., 80, 1083 (2008)
\bibitem{IQHE} von Klitzing K, Dorda G, Pepper M. Phys. Rev. Lett. 45, 494-497, 1980
\bibitem{MR} Gregory W. Moore, N. Read, Nucl. Phys. B 360 (1991) 362-396
\bibitem{Chetan} Chetan Nayak and Frank Wilczek, Nucl. Phys. B 479 (1996) 529-553
\bibitem{Bonderson} Parsa Bonderson, Victor Gurarie and Chetan Nayak, Phys. Rev. B 83, 075303 (2011)
\bibitem{RG} N. Read and Dmitry Green, Phys. Rev. B, 61, 10267 (2000)
\bibitem{Ivanov} D. A. Ivanov, Phys. Rev. Lett. 86, 268 (2001)
\bibitem{salmon} N. B. Kopnin and M. M. Salomaa, Phys. Rev. B, 44, 9667, 1991
\bibitem{volv} G. E. Volovik, JETP Letters, 70, 609, 1999
\bibitem{Alicea_rev} Jason Alicea, Rep. Prog. Phys. 75 (2012) 076501
\bibitem{BdG} P. G. De Gennes, Superconductivity of Metals and Alloys
\bibitem{Lin_Leggett_1} Y. Lin and A. J. Leggett, arXiv:1708.02578
\bibitem{Park} YeJe Park, Suk Bum Chung, and Joseph Maciejko, Phys. Rev. B 91, 054507 (2015)
\bibitem{Lopes} Pedro L. S. Lopes and Pouyan Ghaemi, Phys. Rev. B 92, 064518 (2015)
\bibitem{Sau} Jay D. Sau, B. I. Halperin, K. Flensberg and S. Das Sarma, arXiv: 1106.4014v2, Phys. Rev. B 84, 144509 (2011)
\bibitem{Cheng_1} Meng Cheng and Hong-Hao Tu, arXiv: 1106.2614v2, Phys. Rev. B 84, 094503 (2011)
\bibitem{Cheng_2} Meng Cheng and Roman Lutchyn, arXiv: 1502.04712v2, Phys. Rev. B 92, 134516 (2015)
\bibitem{Iemini} Fernando Iemini, Leonardo Mazza, Davide Rossini, Sebastian Diehl and Rosario Fazio, arXiv: 1504.04230v1
\bibitem{Lang} Nicolai Lang and Hans Peter Buchler, arXiv: 1504.04233v1, Phys. Rev. B 92, 041118 (2015)
\bibitem{Kraus} Christina V. Kraus, Marcello Dalmonte, Mikhail A. Baranov, Andreas M. Lauchli and P. Zoller, arXiv: 1302.0701v1, Phys. Rev. Lett. 111, 173004 (2013)
\bibitem{Ortiz} Gerardo Ortiz, Jorge Dukelsky, Emilio Cobanera, Carlos Esebbag and Carlo Beenakker, arXiv: 1407.3793v1, Phys. Rev. Lett. 113, 267002 (2014)
\bibitem{Alicea_pc} Jason Alicea, to be published
\bibitem{Gurarie} V. Gurarie and L. Radzihovsky, Phys. Rev. B, 75, 212509 (2007)
\bibitem{Ao} Ping Ao and David J. Thouless, Phys. Rev. Lett., 70, 2158 (1993) 
\bibitem{Richardson} R. W. Richardson, Phys. Lett. 3, 277 (1963).
\bibitem{Koch} Jens Koch, Terri M. Yu, Jay Gambetta, A. A. Houck, D. I. Schuster, J. Majer, Alexandre Blais, M. H. Devoret, S. M. Girvin, and R. J. Schoelkopf, Phys. Rev. A 76, 042319, (2007)
\bibitem{Lutchyn} Roman M. Lutchyn, Jay D. Sau, and S. Das Sarma, Phys. Rev. Lett. 105, 077001 (2010)


\bibitem{Anderson} P.W. Anderson, Phys. Rev., 110, 827 (1958)
\bibitem{Anderson_1} P.W. Anderson, Phys. Rev., 112, 1900 (1958)
\bibitem{Feenberg} Eugene Feenberg, American Journal of Physics, 38, 684 (1970)
\bibitem{Leggett_1} A. J. Leggett, J. Phys. C: Solid State Phys., Vol. 6, 1973, 3187-3204
\bibitem{Leggett_2} A. J. Leggett, Ann. Phys. 85, 11-55 (1974)
\bibitem{Leggett_3} A. J. Leggett, Rev. Mod. Phys. Vol. 47, No. 2, 331 (1975)
\bibitem{Silaev} M. A. Silaev, Phys. Rev. B 84, 144508 (2011)
\bibitem{Baym}Gordan Baym and Leo P. Kadanoff, Phys. Rev., 124, 287, 1961; Leo Kadanoff and Paul C. Martin, Phys. Rev., 124, 670, 1961; Gordan Baym, Phys. Rev., 127, 1391, 1962 
\bibitem{Taylor} Edward Taylor, A. John Berlinsky and Catherine Kallin, Phys. Rev. B 91, 134505 (2015)
\bibitem{Kitaev_wire} A Yu Kitaev, arXiv:cond-mat/0010440
\bibitem{WZ} Frank Wilczek and A. Zee, Phys. Rev. Lett. 52, 2111 (1984)
\bibitem{Alicea}Jason Alicea, Yuval Oreg, Gil Refael, Felix Oppen and M.P.A. Fisher, Nature Physics 7, 412 (2011)
\bibitem{Furusaki} Akira Furusaki and Masaru Tsukada, Solid State Communications, 78, 299, 1991
\bibitem{Bagwell} Philip F. Bagwell, Phys. Rev. B, 49, 6841, 1994
\bibitem{Sols} Fernado Sols and Jaime Ferrer, Phys. Rev. B, 49, 15913, 1994

\end{thebibliography}
\end{document}